\documentclass[aps,amsmath,pre,superscriptaddress,showpacs]{revtex4}

\usepackage{graphicx}
\usepackage{epsfig}
\usepackage{amsthm, amsmath, amsfonts, ae}

\begin{document}

\title{A hard-sphere model on generalised Bethe lattices: Dynamics}
 
\author{Hendrik Hansen-Goos} 
\affiliation{Institut f\"ur Theoretische Physik, Universit\"at 
G\"ottingen, Friedrich-Hund-Platz 1, D-37077 G\"ottingen, Germany} 
\author{Martin Weigt} 
\affiliation{Institute for Scientific Interchange, Viale Settimio 
Severo 65, I-10133 Torino, Italy}

\date{\today}

\begin{abstract}
We analyse the dynamics of a hard-sphere lattice gas on generalised
Bethe lattices using a projective approximation scheme (PAS). The
latter consists in mapping the system's dynamics to a finite set of
global observables, closure of the resulting equations is obtained by
approximating the true non-equilibrium state by a pseudo-equilibrium
based only on the value of the observables under consideration. We
study the liquid--crystal as well as the liquid--spin-glass
transitions, special attention is given to the prediction of equilibration
times and their divergence close to the phase transitions. Analytical
results are corroborated by Monte-Carlo simulations.
\end{abstract}
%\pacs{64.70.Pf,64.60.Cn,75.10.Nr}

\maketitle

\section{Introduction}

The statistical mechanics of finite-connectivity systems has seen a
large increase in interest within the last decade. The major reason
for this interest was the emergence of a close interdisciplinary
collaboration between statistical mechanics and theoretical computer
science. In many hard combinatorial problems, phase transitions in
resolvability and algorithmic performance were observed -- once these
problems are suitably randomised \cite{ChKaTa,MiSeLe,KiSe}. This
observation obviously provides a large temptation to use statistical
mechanics tools to get insight into the problem structure going beyond
the one which can be obtained with traditional tools from discrete
mathematics and computer science. The natural mapping of these models
to disordered finite-connectivity models was therefore used to
investigate problems like satisfiability \cite{MoZe,nature,science},
vertex cover \cite{vc_prl}, graph colouring \cite{col_prl} etc.

Coming back to physical models, finite-connectivity systems can be
understood as an intermediate step between fully-connected mean-field
models \cite{MePaVi} and realistic, finite-dimensional systems. Even
if their structure is not geometrical in the sense that they are
embeddable into any finite-dimensional space, the finite connectivity
allows for a reasonable implementation of notations like
neighbourhood, distance etc. This property enables construction and
analysis of microscopically motivated models, in particular for
spin-glasses \cite{MePa,MePa2} and structural glasses
\cite{FrMeRiWeZe,BiMe,weigt,PiTaCaCo,rivoire,HaWe}.

The understanding of these models is, however, mainly based on the
analysis of their thermodynamic equilibrium behaviour. Due to a
technical break-through based on an application of the cavity method
to finite-connectivity models \cite{MePa}, the structure of stable and
metastable states, the appearance of phase transitions etc. are relatively
well-understood by now.

The knowledge is much less evolved in what concerns the dynamics of
finite-connectivity models
\cite{BaZe,MoRi,CuSe,SeCuMo,SeWe,coolen1}. Here we will use an
approximation scheme first introduced to analyse the dynamical behaviour of
stochastic local-search algorithms for combinatorial optimisation
\cite{SeMo,BaHaWe}. We will, however, use a different formulation
which is based on the observation from Ref.~\cite{SeWe} that the method can be
put into the canonical form of a projective approximations scheme
being equivalent to the dynamical replica theory
\cite{CoSh,CoSh2,LaCoSh} which was originally developed for fully
connected systems. Note also \cite{coolen2}, where the same method was
recently applied to Ising models on random graphs.

The model studied here is a hard-sphere lattice-gas model on a
generalised Bethe lattice, which was first proposed in \cite{weigt} as
a microscopically motivated, but solvable model for the structural
glass transition. The equilibrium behaviour of this model, including
crystallisation as well as glass formation, was studied in great
detail in \cite{HaWe}. The present work, which is concentrated on the
relaxational dynamics of the model, can be seen as a natural
continuation of the work presented therein. Even if we have tried to
make the present paper as self-contained as possible, sometimes we
have to refer to technical details in the previous publication --
whose full inclusion would go beyond the scope of the present article.

The paper is organised as follows. Sec.~\ref{sec:model} gives an
introduction to the system including a brief review of its static
properties and the definition of the dynamics. In the following
section, the general outline of the applied approximation scheme is
presented. Sec.~IV is dedicated to crystallisation dynamics,
containing two approximations and comparison with Monte-Carlo simulations. In
Sec.~V an approximation of the dynamics near the spin-glass transition
is presented. Finally, conclusion and outlook are given in the last
section.

\section{The model}
\label{sec:model}

In a first step, we are going to define the model itself, review
its static properties which were analysed in detail in \cite{HaWe}. In
the second part of this section we will define microscopic physical
dynamics for the model, whose analysis will be the subject of the whole
paper.

\subsection{Construction and static properties}

\begin{figure}[tbp]
  \begin{center}
    \includegraphics[height = 6cm]{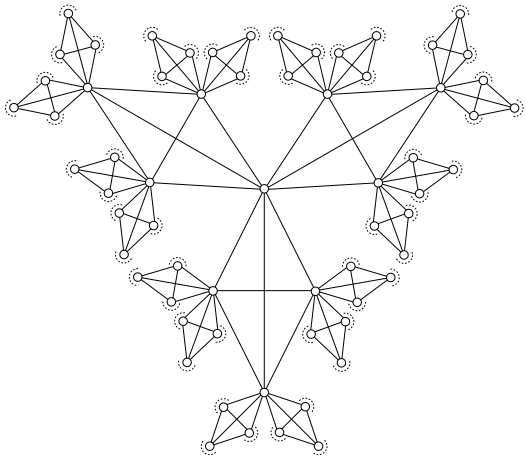}
    \caption{Part of a generalised Bethe-lattice with $k=2$,
    $p=3$. The 21 cliques connecting the central vertex to its
    nearest and second neighbours are shown. Figure from Ref.~\cite{HaWe}.}
\label{fig_allgbethe}
  \end{center}
\end{figure}

The system is a lattice gas of hard spheres living on a generalised
Bethe lattice. Let us first specify the character of the
particles. The property of hard spheres is realised by introducing an
excluded volume associated with each particle. This is done by requiring
that two neighbouring sites (i.e. sites that are connected with an edge
in the graph) cannot be occupied simultaneously. Two particles have
thus a minimal distance of two on the lattice.

The generalised Bethe lattice is constructed in the following way: The
basic building blocks are small completely connected sub-graphs
(i.e. cliques), each of which contains $p+1$ vertices. In each vertex
$k+1$ of these cliques are merged. This procedure leads to a graph
which is locally isomorphic to that shown in Fig.~\ref{fig_allgbethe}
for one specific choice of $k$ and $p$. The ideal graph can be thought
of as an infinite continuation of the shown structure such that the
whole graph remains loop-less apart from the inner-clique loops. Note
that the ordinary Bethe lattice can be obtained as the special case
$p=1$. It is furtheron obvious that a close packing on such a lattice
consists of a configuration where each clique carries exactly one
particle -- which, because of the hard-sphere constraint, realises also
the maximum particle density that can be found in a single clique.

A treatment of a lattice gas on this graph would, however, lead to
subtle questions about the boundary conditions. In the thermodynamic
limit, we would face a strange situation where the boundary is
infinitely far from a fixed centre, but the number of boundary sites
constitute a finite fraction of all sites. As we shall recur to finite
lattices when confronting our results with Monte-Carlo data,
it seems more appropriate to base the calculations on the properties
of finite random realisations of the generalised Bethe lattice from
the beginning, see also the discussion in \cite{HaWe}.

A finite graph allowing for a close packing as described above can be
constructed, e.g., as a bipartite (or $(p+1)$-partite) graph which
possesses one (or $p+1$) close packings. Any given close packing defines two
sub-lattices referred to as 0-lattice (empty sites) and 1-lattice
(occupied sites) in the following. The equilibrium properties of the
system can be obtained from a calculation of the grand canonical
partition function
\begin{equation}
\label{eq:Xi}
  \Xi(\mu) = \sum_{n_1, \ldots , n_N \in \{0, 1\}}
  e^{\mu\sum_{i=1}^{N}n_i}\,\prod_{\{i,j\}\in E} (1-n_i n_j)
\end{equation}
where $\mu$ is the chemical potential, $i, j = 1, \ldots , N$ label
the sites and $E$ is the set of edges of the graph. The calculation
was carried out in detail in \cite{HaWe}. One finds that the system
has a homogeneous low-$\mu$-phase (i.e.~both sublattices have same
particle densities) which is referred to as the {\em liquid} phase. In
the high-$\mu$-regime the system can be found in a phase where the
1-lattice has a higher particle density than the 0-lattice. This phase
is called {\em crystalline}. In the case $p>1$ there is also an
intermediate regime where both liquid and crystalline phases are
locally stable, and the phase transition between liquid and crystal
is of first order. This explains the interest of generalising the
concept of an ordinary Bethe-lattice ($p=1$) which lacks the
intermediate phase. The extension can be motivated physically by the
abundance of short loops found in finite dimensional lattices. They
are taken into account at least approximately by the introduction of
local cliques.

Another possibility to realise the finite graph is to construct a
generalised regular random graph, i.e. a graph that is locally
isomorphic to a generalised Bethe lattice but contains long loops with
a typical length that is logarithmic in $N$. This setting leads to a
qualitatively different high-$\mu$-phase: The graph is, with high
probability, no longer compatible with a crystalline close
packing. The statics of the system can be appropriately calculated via
the cavity approach, cf. \cite{MePa,rivoire}. The scenario for
arbitrary $k$ and $p$ is described in \cite{HaWe}. In the present
study we only focus on the case where $p=1$ (ordinary
Bethe lattice). The system then displays a liquid low-$\mu$-phase and
a continuous transition to an amorphous high-$\mu$ spin-glass phase. A
one-step-replica-symmetry-broken (1RSB) ansatz takes into account the
frustration due to odd loops, and gives a lower grand-potential than
the liquid solution.

\subsection{Definition of the dynamics}
\label{sec_defdyn}

The dynamics of the model is made up by two basic processes: The first
one being diffusion of particles on the lattice, the second one
particle exchange with an infinite particle bath. The realisation of
these processes is as follows:
\begin{itemize}
\item In every time step, a lattice site is selected randomly.
\item If the lattice site is occupied, with probability $q_0$ the
particle is annihilated (or, equivalently, transferred to the particle
bath). With probability $q_s$, the particle tries to jump to a
randomly selected neighbouring lattice site. This jump is only
realisable, if the new lattice site is empty itself, and has no
further occupied neighbouring sites. Every other case would either
result in two particles on one site, or in two neighbouring particles,
which is excluded according to Eq.~(\ref{eq:Xi}).
\item If the lattice site chosen in the first step is empty, we try to
create a new particle with probability $q_1$. This trial is
successful only if also all neighbouring sites of the selected one
are empty.
\end{itemize}
The Monte-Carlo (MC) dynamics defined in this way approaches the
thermodynamic equilibrium described by Eq.~(\ref{eq:Xi}) if the rates
for particle annihilation and particle creation are related by
detailed balance:
\begin{equation}
\label{eq:detailled}
e^\mu = \frac {q_1}{q_0}
\end{equation}
which we assume to be satisfied throughout this paper. Besides the
fact that the described dynamics can be easily implemented on
arbitrary finite lattices or graphs, it can also be understood as a
valid microscopic dynamics of the physical model under consideration.

The time unit is one MC sweep ($\Delta t=1$) which corresponds to a
number $N$ of the above trials. A single trial thus corresponds to
$\Delta t = 1/N$, and the time variable becomes continuous in the
thermodynamic limit.

\section{Projective approximation schemes}
\label{sec:schemes}

The dynamics of diluted systems have been described successfully in
the past by use of rate equations \cite{SeMo,BaHaWe,SeWe,coolen2}.
This method, even if formulated differently in some of these
references, has an elegant formalisation as a {\it projective
approximation scheme} (PAS) described in the following steps:
\begin{itemize}
\item A set of intensive macroscopic observables $\vec{\mathcal{O}}$
  is chosen. Typically $\vec{\mathcal{O}}$ are densities and we
  require ``$\vec{\mathcal{O}}\supseteq \vec{\mathcal{O}}_{equil.}$'',
  i.e. the set must contain the observables necessary for the
  discrimination of the different phases at equilibrium (e.g. the two
  sublattice densities to describe crystallisation in our model).
\item In a general non-equilibrium situation, the dynamical equations
  for the observable do not close in the set $\vec{\mathcal{O}}$. One
  can, however, approximately close the equations by assuming
  equilibrium in a generalised ensemble: All microscopic
  configurations leading to the same values of all observables in
  $\vec{\mathcal{O}}$ are assumed to be equiprobable.
\end{itemize}
There is, of course, no reason why this closure assumption should be
true in general. It can, however, be seen as the most natural
approximation to the actual distribution of the system which is based
solely on the knowledge of the observable values, and on no further
information. In this sense, the PAS has to be considered a the best
possible approximation of the model's dynamics based on $\vec{\mathcal{O}}$. This does not mean that,
for specific initial conditions, parameter values etc., there cannot
be a better approximation to the actual time dependence of the
observables. For the general case, it constitutes, however, an optimal
use of the available information on the system. Further on, since we
required ``$\vec{\mathcal{O}}\supseteq \vec{\mathcal{O}}_{equil.}$'',
we are sure that the true thermal equilibrium is a fixed point of the
PAS.

Technically, the equipartition hypothesis is realised in the following
way:
\begin{itemize}
\item For each observable in $\vec{\mathcal{O}}$, a conjugate chemical
  potential is introduced (e.g. the usual chemical potential $\mu$
  would couple to the particle density $\rho$). The potentials are
  collected in the vector $\vec{\mu}$.
\item The generalised grand partition function 
  $$ 
  \Xi_G(\vec \mu) = \sum_{\{n_i\}} e^{N\
  \vec{\mu}\cdot\vec{\mathcal{O}}} \,\prod_{\{i,j\}\in E} (1-n_i n_j)
  $$ 
  automatically implements the equipartition condition.
\item For given observable values of $\vec{\mathcal{O}}$, the values
  of the chemical potentials $\vec{\mu}$ are obtained
  self-consistently from $\Xi_G(\vec \mu)$.
\item Starting from the values of the observables $\vec{\mathcal{O}}$,
  we derive equations for the time changes
  $\dot{\vec{\mathcal{O}}}$. The latter depend in general on a larger
  set of observables, whose values are evaluated in the generalised
  equilibrium ensemble using the previously determined chemical
  potentials. In the next chapter it will become clear, how this step
  is realised technically.
\end{itemize}
This scheme provides a well-defined prescription for the approximation
of the dynamics of the system based on the observable set
$\vec{\mathcal{O}}$. Unfortunately this prescription does not include
any intrinsic way of quantifying deviations from equipartition,
such that all results based on this scheme have to be cross-checked
against numerical simulations. On the other hand, it is obvious that
going to more detailed observables means to enhance the quality of the
approximation, at the cost of introducing a more complex generalised
ensemble. This will be shown in detail for the crystallisation
dynamics below in Sec.~\ref{sec:crystal}.

We mention that the assumption of equipartition of probability was
introduced in the context of dynamical replica theory \cite{CoSh} for
fully connected spin-glass models. In this context, the technical
treatment of $\Xi_G(\vec\mu)$ was obtained using the replica trick. In
the context of finite-connectivity systems, however, the application
of the cavity method will be much more efficient.

\section{Crystallisation dynamics}
\label{sec:crystal}

As a first application of PAS to our model, we are going to study the
crystallisation dynamics of the model. We will show two levels of PAS,
the first one being the minimal description leading to a correct
characterisation of the equilibrium, the second one will include a
larger set of observables. Both schemes will be compared to numerical
data obtained from MC simulations on large, but finite lattices. The
experience gained in this context will be used in a later section in
order to approximate also the dynamics close to the spin-glass
transition.

\subsection{The $\rho$-approximation}

\subsubsection*{From the generalised ensemble to approximate dynamical equations}
\label{sec_rhoapprox}

\begin{figure}[tbp]
  \begin{center}
    \includegraphics[height = 3.7cm]{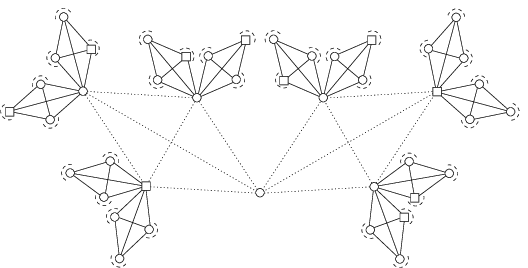}
    \hspace{0.4cm}
    \includegraphics[height = 3.7cm]{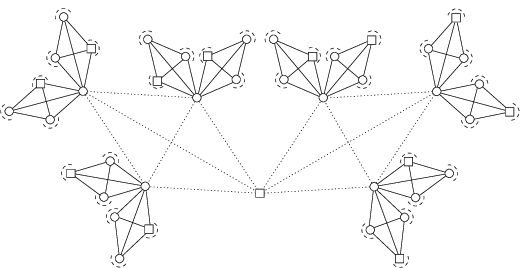}
    \caption{Possible iteration steps leading to rooted trees with
      root in the 0-lattice (left), 1-lattice (right). Vertices
      belonging to the 0-lattice (1-lattice) are depicted by $\circ$
      ($\Box$). Figure from Ref.~\cite{HaWe}.}
\label{fig_ocemiter}
  \end{center}
\end{figure}

Let us start with a minimal set of observables. As we need an
indication whether the system is liquid or crystalline, we obviously
need the two sublattice particle densities $\rho^{(0)}$ and
$\rho^{(1)}$ that refer to the 0-lattice and 1-lattice
respectively. The normalisation is chosen such that $\rho^{(0)}\to 0$
and $\rho^{(1)} \to 1$ in the close packing limit $\mu\to\infty$.
Before being able to write any closed dynamical equation for the two
sublattice densities, we have to consider the generalised equilibrium
ensemble defined in the last section. We introduce two chemical
potentials $\mu^{(0)}$ and $\mu^{(1)}$ that couple to the particles on
the respective sublattice. The calculation of $\Xi_G$ can be performed
analogously to that of the grand partition function $\Xi(\mu)$ using
the concept of rooted trees. Here we give only a shortened description
of how to do so, for technical details see \cite{HaWe}. A rooted tree
is obtained from a generalised Bethe lattice by choosing a site as the
root and removing one of the $k+1$ incident branches. Rooted trees can
be iterated according to Fig.~\ref{fig_ocemiter} where dashed lines
depict the edges which are added in one iteration step. We introduce
conditional partition functions $\Xi_e^{(0/1)}$
(resp. $\Xi_*^{(0/1)}$) for a subtree rooted in a 0/1-site which is
fixed to be empty (resp. occupied). These quantities iterate according
to
\begin{subequations}
\label{gl_rhoApproxIt}
\renewcommand{\theequation}{\theparentequation \roman{equation}}
\begin{eqnarray}
 {\Xi_e^{(0)}}' & = & \Bigl (\Xi_e^{(0)}\bigl(\Xi_e^{(0)}\bigr)^{p-1} +
    \Xi_*^{(1)}\bigl(\Xi_e^{(0)}\bigr)^{p-1}+
    (p-1)\,\Xi_e^{(1)}\Xi_*^{(0)}\bigl(\Xi_e^{(0)}\bigr)^{p-2} \Bigr)^k
    \\
  {\Xi_*^{(0)}}' & = &
  e^{\mu^{(0)}}\,\Bigl(\Xi_e^{(1)}\bigl(\Xi_e^{(0)}\bigr)^{p-1}
  \Bigr)^k \\
  {\Xi_e^{(1)}}' & = & \Bigl(\bigl(\Xi_e^{(0)}\bigr)^{p} +
  p\,\Xi_*^{(0)}\bigl(\Xi_e^{(0)}\bigr)^{p-1}  \Bigr)^k \\
  {\Xi_*^{(1)}}' & = &
  e^{\mu^{(1)}}\,\Bigl(\bigl(\Xi_e^{(0)}\bigr)^p\Bigr)^k \; .
\end{eqnarray}
\end{subequations}

The stroke labels the partition functions of the resulting large
rooted tree. All 0-rooted trees that enter in the iteration are taken
to be identical, the same is assumed for all 1-rooted trees. When we
pass to intensive quantities by introducing local fields $x^{(0/1)} =
\frac{\Xi_*^{(0/1)}}{\Xi_e^{(0/1)}}$, the strokes in
Eqs.~\eqref{gl_rhoApproxIt} drop out and the iteration equations
reduce to two equations for the local fields $x^{(0/1)}$ which still
contain the unknown chemical potentials $\mu^{(0/1)}$. These can be
eliminated through the connection of Eqs.~\eqref{gl_rhoApproxIt} with
the observables $\rho^{(0/1)}$ which is established via the physical
partition functions $\tilde{\Xi}_e^{(0/1)}$ and
$\tilde{\Xi}_*^{(0/1)}$. The latter refer to a site within the
generalised Bethe lattice (not the root of a subtree), i.e.~they are
obtained like the stroked partition functions in
Eqs.~\eqref{gl_rhoApproxIt} after the substitution $k \to k+1$ on the
r.h.s. of the equations. The relations
$\frac{\rho^{(0/1)}}{1-\rho^{(0/1)}} =
\frac{\tilde{\Xi}_*^{(0/1)}}{\tilde{\Xi}_e^{(0/1)}}$ allow to derive
two more equations, and we end up with four equations containing the
four unknown $x^{(0/1)}$ and $\mu^{(0/1)}$. Their solution is
straight-forward yielding
\begin{equation}
  x^{(0/1)} = \frac{\rho^{(0/1)}}{1-\rho^{(1)}-p\,\rho^{(0)}}
  \label{gl_x01approx}
\end{equation} 
and
\begin{equation}
  e^{\mu^{(0/1)}} =
  \frac{\rho^{(0/1)}(1-\rho^{(0/1)})^k}
  {(1-\rho^{(1)}-p\,\rho^{(0)})^{k+1}} \; . \label{gl_mu01}
\end{equation}
We note that Eq.~(\ref{gl_mu01}) is the self-consistency result for
the chemical potentials and Eq.~\ref{gl_x01approx} is
the result for the conditional partition functions that are
essentially contained in the local fields. We
also mention that by setting $\mu^{(0)}=\mu^{(1)}=\mu$ the results of
the calculation for the statics are recovered, i.e. we have found an
``equilibrium'' treatment for non-equilibrium configurations of
$\rho^{(0/1)}$ that naturally contains the physical equilibrium
ensemble as a special case.

In order to derive the approximative differential equations for the
time-evolution of $\rho^{(0/1)}$, we start with the exact equations
for the average change $\Delta N_*^{(0/1)}$ in the number of particles
$N_*^{(0/1)}$ on the respective sublattice. The average is taken over
the time $\Delta t = 1/N$ where $N = N^{(0)}+N^{(1)}$ is the overall
number of sites and $N^{(0/1)}$ are the numbers of sites belonging to
the sublattices. $\Delta t$ is the time that corresponds to one action
of the MC-algorithm that implements the dynamics (see
Sec.~\ref{sec_defdyn}). The changes $\Delta N_*^{(0/1)}$ are given by
rate-equations:
\begin{eqnarray}
  \Delta N_*^{(0/1)} & = & -\frac{N^{(0/1)}}{N} \, \rho^{(0/1)} \, q_0
    + \frac{N^{(0/1)}}{N} \, (1-\rho^{(0/1)}) \, q_1 \,
    \mathcal{P}_1^{(0/1)} \nonumber \\
    & & - \frac{N^{(0/1)}}{N} \, \rho^{(0/1)} \, q_s \,
    \mathcal{P}_s^{(0/1)} 
    + \frac{N^{(1/0)}}{N} \, \rho^{(1/0)} \, q_s \,
    \mathcal{P}_s^{(1/0)} \; . \label{gl_ratenrho}
\end{eqnarray}

The first term on the r.h.s. corresponds to a loss due to annihilation
of a particle. The three factors mean that the algorithm must choose a
site on the lattice in question ($\frac{N^{(0/1)}}{N}$) which is
occupied ($\rho^{(0/1)}$) and annihilate it ($q_0$). This contribution
is exactly known. The second term contains the gain due to particle
creation. Here $\mathcal{P}_1^{(0/1)}$ denotes the probability that
the particle creation is possible without violating the hard-sphere
constraint, i.e.~the probability that all neighbours of the selected
site on the 0/1-lattice are also empty. Its value in one microscopic
(non-equilibrium) configuration cannot be expressed as a function of
the $\rho^{(0/1)}$-values of this configuration, as it contains
information about nearest-neighbour correlations that are not included
in the single-site observables $\rho^{(0/1)}$. The third term gives
the loss of particles caused by those that jump to the other
sublattice whereas the last term contains the gain of particles caused
by those that jump from the other sublattice to the sublattice under
consideration. Hereby, $\mathcal{P}_s^{(0/1)}$ denotes the probability
that a tried move of a particle on the 0/1-lattice to a neighbouring
site on the 1/0-lattice is in fact feasible. This is again a quantity
which involves neighbour correlations and cannot be expressed exactly
for one configuration given only by $\rho^{(0/1)}$.

In order to close Eq.~\eqref{gl_ratenrho} we now proceed to the last
step mentioned in the outline of PAS. We have to calculate the
probabilities $\mathcal{P}_{\cdot}^{(0/1)}$ within the pseudo-equilibrium
ensemble characterised by the already determined formal chemical
potentials $\mu^{(0/1)}$, i.e.~the $\mathcal{P}_{\cdot}^{(0/1)}$ are
calculated via a flat average over all configurations having the
desired values of $\rho^{(0/1)}$. For instance, we can calculate
\begin{eqnarray}\label{gl_P10rechnung}
  \mathcal{P}_1^{(0)} & = &
  \frac{\overbrace{
      \Bigl(\Xi_e^{(1)}\bigl(\Xi_e^{(0)}\bigr)^{p-1}\Bigr)^{k+1}}
    ^{\text{all neighbouring sites are empty}}} 
       {\underbrace{ \Bigl (\Xi_e^{(1)}\bigl(\Xi_e^{(0)}\bigr)^{p-1} +
           \Xi_*^{(1)}\bigl(\Xi_e^{(0)}\bigr)^{p-1}+
           (p-1)\,\Xi_e^{(1)}\Xi_*^{(0)}\bigl(\Xi_e^{(0)}\bigr)^{p-2} 
     \Bigr)^{k+1} }_{\text{no requirements to the neighbouring sites}}} 
       \nonumber \\
  & = & \frac{1}{\bigl(1+x^{(1)}+(p-1)\,x^{(0)}\bigr)^{k+1}} \nonumber
  \\
  & \overset{\text{\eqref{gl_x01approx}}}{=} & \left
    (\frac{1-\rho^{(1)}-p\,\rho^{(0)}} {1-\rho^{(0)}} \right)^{k+1} \; .
\end{eqnarray}
Similarly we find
\begin{equation}
  \mathcal{P}_1^{(1)} =  \left
    (\frac{1-\rho^{(1)}-p\,\rho^{(0)}} {1-\rho^{(1)}} \right)^{k+1} \; .
\end{equation}

For the calculation of $\mathcal{P}_s^{(0)}$ we must consider that,
starting from a 0-site, only $k+1$ of the $(k+1)p$ neighbouring sites
belong to the 1-lattice which is taken into account with a factor
$\frac{1}{p}$. Furthermore we know that the ending site of the jump is
empty as are all other sites in the clique that is shared by starting
site and ending site. The only condition for $\mathcal{P}_s^{(0)}$ is that
the $kp$ sites of the $k$ other cliques containing the ending site
are empty. These sites are all in the 0-lattice as the ending site is
in the 1-lattice. This leads to
\begin{equation}
   \mathcal{P}_s^{(0)} = \frac{1}{p} \: \frac
   {\Bigl(\bigl(\Xi_e^{(0)}\bigr)^{p}\Bigr)^k} {
     \Bigl(\bigl(\Xi_e^{(0)}\bigr)^{p} +
     p\,\Xi_*^{(0)}\bigl(\Xi_e^{(0)}\bigr)^{p-1}  \Bigr)^k}
  =  \frac{1}{p} \: \left
    (\frac{1-\rho^{(1)}-p\,\rho^{(0)}} {1-\rho^{(1)}} \right)^{k}
\end{equation}
and similarly
\begin{equation}
  \mathcal{P}_s^{(1)} =  \left
    (\frac{1-\rho^{(1)}-p\,\rho^{(0)}} {1-\rho^{(0)}} \right)^{k} \; .
\end{equation}

These approximations close Eq.~\eqref{gl_ratenrho}. We add that
\begin{equation}
  \Delta N_*^{(0/1)} = \frac{N^{(0/1)}}{N}\,\frac{\Delta
    (N_*^{(0/1)}/N^{(0/1)})}{\Delta t} = \frac{N^{(0/1)}}{N}\,\frac{\Delta
    \rho^{(0/1)}}{\Delta t}
\end{equation}
and
\begin{equation}
  \frac{N^{(0)}}{N} = \frac{p}{p+1}\, , \qquad  
  \frac{N^{(1)}}{N} = \frac{1}{p+1}\; .
\end{equation}
for the generalised Bethe lattice.

Using these ingredients, in the
thermodynamic limit $N\to\infty$ Eq.~\eqref{gl_ratenrho} becomes:
\begin{subequations}
\label{gl_rhoapprox}
\renewcommand{\theequation}{\theparentequation \roman{equation}}
\begin{align}
  \dot{\rho}^{(0)}  = & - \rho^{(0)} \, q_0 + (1-\rho^{(0)})\, q_1 \,
  \left(\frac{1-\rho^{(1)}-p\,\rho^{(0)}}{1-\rho^{(0)}}\right)^{k+1}
  \nonumber \\
  &  -\rho^{(0)}\,q_s\,\frac{1}{p}\, 
    \left(\frac{1-\rho^{(1)}-p\,\rho^{(0)}}{1-\rho^{(1)}}\right)^{k} 
  +\rho^{(1)}\,q_s\,\frac{1}{p}\, 
  \left(\frac{1-\rho^{(1)}-p\,\rho^{(0)}}{1-\rho^{(0)}}\right)^{k}
 \label{gl_rho0approx}\\
  \intertext{and}
  \dot{\rho}^{(1)}  = & - \rho^{(1)} \, q_0 + (1-\rho^{(1)})\, q_1 \,
  \left(\frac{1-\rho^{(1)}-p\,\rho^{(0)}}{1-\rho^{(1)}}\right)^{k+1}
  \nonumber \\
  &  -\rho^{(1)}\,q_s\, 
    \left(\frac{1-\rho^{(1)}-p\,\rho^{(0)}}{1-\rho^{(0)}}\right)^{k} 
  +\rho^{(0)}\,q_s\, 
  \left(\frac{1-\rho^{(1)}-p\,\rho^{(0)}}{1-\rho^{(1)}}\right)^{k} \,
  . \label{gl_rho1approx}
\end{align}
\end{subequations}
Eqs.~\eqref{gl_rhoapprox} constitute what we call the
$\rho$-approximation, i.e.~the approximation for the crystallisation
dynamics which is derived for a minimal set of observables --
$\rho^{(0)}$ and $\rho^{(1)}$ -- in the context of PAS.
The approximative time-flow of $\rho^{(0/1)}$ can be obtained for
given initial conditions by integrating Eqs.~\eqref{gl_rhoapprox}
numerically. Results are shown in Sec.~\ref{sec:MCcrystal} where the
$\rho$-approximation and $\sigma_j$-approximation are compared with
MC simulations.

\subsubsection*{Asymptotics of the $\rho$-approximation}
\label{sec:rhoasymp}

Exactly at the stationary point of these equations, corresponding to
an equilibrium solution of the model, the assumption of the PAS
becomes true. The stationary points of Eqs.~\eqref{gl_rhoapprox} are
therefore identical to the static liquid and crystalline solutions
discussed in Ref.~\cite{HaWe}.

A linear expansion of Eqs.~(\ref{gl_rhoapprox}) therefore allows to
analyse the stability of the equilibrium solution as well as to
determine the equilibration time. To achieve this, we have to
calculate the Jacobian of the above system of equations,
\begin{equation}
  A \overset{.}{=} \left( 
        \begin{array}{cc}
           \frac{\partial\dot{\rho}^{(0)}}{\partial\rho^{(0)}} &  
           \frac{\partial\dot{\rho}^{(0)}}{\partial\rho^{(1)}} \\
           \frac{\partial\dot{\rho}^{(1)}}{\partial\rho^{(0)}} &  
           \frac{\partial\dot{\rho}^{(1)}}{\partial\rho^{(1)}}
        \end{array}
      \right) \ .
\end{equation}
This matrix has to be evaluated at the stationary point. Let us assume
that it has the two eigenvalues $\lambda_1$ and $\lambda_2$ with corresponding eigenvectors $\vec v_1$ and $\vec v_2$. The
time evolution of a small perturbation around the stationary point is
then given by
\begin{equation}\label{gl_algloesasymp}
   \Biggl( 
        \begin{array}{c}
           \rho^{(0)}(t) \\
           \rho^{(1)}(t)
        \end{array}
      \Biggr)
      =   
    \Biggl( 
        \begin{array}{c}
           \rho^{(0)}_{\text{stat}} \\
           \rho^{(1)}_{\text{stat}}
        \end{array}
      \Biggr)
      + \alpha_1\,e^{\lambda_1 t}\,\vec{v}_1
      + \alpha_2\,e^{\lambda_2 t}\,\vec{v}_2
\end{equation} 
with constants $\alpha_1$ and $\alpha_2$ which follow from the initial
condition, i.e. from the specific form of the perturbation.

The stationary point is obviously stable if both eigenvalues
$\lambda_1$ and $\lambda_2$ are negative. In this case, the
equilibration time is given by $\tau_{eq} =
-1/\max\{\lambda_1,\lambda_2\}$. If at least one of the eigenvectors
is positive, the stationary point is unstable and thus physically
irrelevant: Any small perturbation having a component in the direction
of the corresponding eigenvector carries the system exponentially fast
away from the stationary point.

Let us first consider the {\it liquid solution}, which is
characterised by equal sublattice densities
$\rho^{(0)}=\rho^{(1)}=\rho$. In this case, the eigenvalues can be
given in a compact form:
\begin{subequations}
\label{gl_eigenliquid}
\renewcommand{\theequation}{\theparentequation \roman{equation}}
\begin{eqnarray}
  \lambda_1 & = & - \frac{q_1\bigl(1+\rho(kp-1)\bigr)}
  {e^{\mu}\bigl(1-(p+1)\rho\bigr)\bigl(1-\rho\bigr)}  \\
  \lambda_2 & = & - \frac{\bigl(1-(k+1)\rho\bigr)\bigl(q_1 p + \rho
    (p+1) (q_s-q_1 p) \bigr)}
  {e^{\mu}\,p\,\bigl(1-(p+1)\rho\bigr)\bigl(1-\rho\bigr)} \,
\end{eqnarray}
\end{subequations}
and the eigenvectors are
\begin{equation} \label{gl_eigenvecsfl}
  \vec{v}_1 =   \biggl( 
        \begin{array}{c}
           1 \\
           1
        \end{array}
      \biggr) \, , \qquad\qquad 
  \vec{v}_2 =   \biggl( 
        \begin{array}{c}
           1 \\
           -p
        \end{array}
      \biggr) \, .
\end{equation}
Both eigenvectors are negative for low densities, but $\lambda_2$
changes sign at $\rho_{cr} = \frac{1}{k+1}$. This point coincides with
the local instability of the liquid state in the statical calculation
of \cite{HaWe}.

It is also interesting to look at the equilibration time $\tau_{eq} =
-1/\max\{\lambda_1,\lambda_2\}$. For small chemical potential $\mu$,
the latter is given by $-1/\lambda_1$. Looking at the eigenvector
$\vec v_1$, we see that the corresponding slowest equilibration
process in the system describes the decay of global density
fluctuations. Close to the instability point of the liquid solution,
we find, however, $\tau_{eq} =-1/\lambda_2$, and the eigenvector
indicates that the slowest decaying perturbation is a difference in
the two sublattice densities, i.e. a fluctuation toward a crystalline
state. At a certain $\mu=\mu_{eq}$ the two eigenvectors are equal. At
this point, a crossover between the two different slowest processes
takes place. It is also the point, where the equilibration time takes
its minimum inside the liquid phase, i.e. where the system
equilibrates faster than for any other $\mu<\mu_{cr}$, see also the
discussion in Sec.~\ref{sec:MCcrystal}, where these analytical results
are compared with MC simulations.

Let us now discuss the stability of the two {\it crystalline
solutions} calculated in \cite{HaWe}. The determinant of the Jacobian matrix $A$ vanishes at the
point where the crystalline solution appears.
\begin{itemize}
\item
For $p=1$, i.e. on the ordinary Bethe lattice, the two crystalline
solutions appear continuously out of the liquid solution, and they
become both stable, marking thus a second-order transition. The two
solutions are in fact identical, they are related by an exchange of
the two sub-lattices which are isomorphic for $p=1$.
\item
For $p>1$, the two crystalline solutions appear discontinuously. The
normal crystalline solution, i.e. the one with higher total density,
becomes immediately locally stable, both eigenvalues of the Jacobian
matrix are negative.  The second solution is first locally unstable
and thus unphysical. At $\mu=\mu_{cr}$, the two sublattice densities
cross, however, at the value $\rho_{cr} = \frac{1}{k+1}$, and thus
coincide with the liquid solution. For higher $\mu$ we have
$\rho_0>\rho_1$, in \cite{HaWe} we have introduced the name ``inverse
crystallisation'' for this phenomenon. This solution takes over the
local stability of the liquid solution as can be inferred from the
eigenvalues of $A$.
\end{itemize}

It is also possible to analytically characterise the critical behaviour
close to $\mu_{cr}$, which is, on both sides of the transition,
characterised by a critical exponent $-1$:
\begin{equation}
\tau_{eq} \propto | \mu-\mu_{cr} |^{-1}\ .
\end{equation}
We skip the calculation of the prefactors here, we only mention that
in the case $p=1$
both sides of the transition are related by an universal factor $1/2$,
as already found in the case of a simple ferromagnet on a Bethe
lattice \cite{SeWe}. Further results are given below in the context of
the comparison between analytical and numerical data.

\subsection{The $\sigma_j$-approximation}
\label{sec_sigapprox}

The $\rho$-approximation was derived according to PAS using the pure
single-site observables $\rho^{(0/1)}$. There we have seen that it was
necessary to approximate nearest-neighbour correlations within the
generalised ensemble. We now go over to more detailed observables
including also such nearest-neighbour correlations. This can be
achieved by defining
\begin{equation}
\begin{tabular}{lp{9.5cm}}
  $\sigma_j^{(0/1)}$:= & density of sites on the 0/1-lattice which are
  empty and have exactly $j$ occupied neighbouring sites, 
  $j=0, \ldots , k+1$.
\end{tabular} 
\end{equation} 
Note that the $\sigma_j^{(0/1)}$ imply the previous observables via
$\rho^{(0/1)} = 1 - \sum_{j=0}^{k+1} \sigma_j^{(0/1)}$. Therefore they
allow for an exact characterisation of the liquid and crystalline
equilibrium states, and according to the rule of our PAS they
represent a valid set of observables.

Again, we apply PAS introducing formal chemical potentials
$\mu_j^{(0/1)}$, $j = 0, \ldots , k+1$, being conjugated to the
$\sigma_j^{(0/1)}$. This allows to calculate conditional partition
functions $\Xi_j^{(0/1)}$, $j= 0, \ldots , k$ which stand for a
0/1-rooted tree with an empty root having $j$ occupied neighbouring
sites, and $\Xi_*^{(0/1)}$ which stands for a 0/1-rooted tree with
occupied root.  The iteration of rooted trees (see
Fig.~\ref{fig_ocemiter}) reads in terms of the conditional partition
functions:
\begin{subequations}
\label{gl_sigmarek}
\renewcommand{\theequation}{\theparentequation \roman{equation}}
\begin{eqnarray}
{\Xi_*^{(0)}}' & = &
\Biggl\{\biggl(\sum_{l=0}^{k}e^{\mu_{l+1}^{(1)}}
    e^{-\mu_{l}^{(1)}}\Xi_l^{(1)}\biggr)
  \biggl(\sum_{l=0}^{k}e^{\mu_{l+1}^{(0)}}
    e^{-\mu_{l}^{(0)}}\Xi_l^{(0)}  \biggr)^{p-1} \Biggr\}^k \, ,
    \label{gl_sigmarek1} \\
{\Xi_j^{(0)}}' & = & e^{\mu_j^{(0)}} {k \choose j} 
  \Biggl\{ \Xi_*^{(1)} \biggl( \sum_{l=0}^{k} e^{\mu_{l+1}^{(0)}}
    e^{-\mu_{l}^{(0)}} \Xi_l^{(0)}\biggr)^{p-1} \nonumber \\ 
& &  \hspace{2cm} + (p-1)\,\Xi_*^{(0)}
  \biggl( \sum_{l=0}^{k} e^{\mu_{l+1}^{(1)}} e^{-\mu_{l}^{(1)}}
    \Xi_l^{(1)}\biggr) \biggl( \sum_{l=0}^{k} e^{\mu_{l+1}^{(0)}}
    e^{-\mu_{l}^{(0)}} \Xi_l^{(0)}\biggr)^{p-2} \Biggr\}^j \times \nonumber 
\\ & & \hspace{1.18cm}\times\:\Biggl\{\biggl(\sum_{l=0}^{k}\Xi_l^{(1)} \biggr)
  \biggl(\sum_{l=0}^{k}\Xi_l^{(0)}\biggr)^{p-1} \Biggr\}^{k-j}
  \label{gl_sigmarek2} \\
  \text{and} \nonumber \\
{\Xi_*^{(1)}}' & = & \Biggl\{\biggl(\sum_{l=0}^{k}e^{\mu_{l+1}^{(0)}}
    e^{-\mu_{l}^{(0)}}\Xi_l^{(0)}  \biggr)^{p} \Biggr\}^k \, ,
    \label{gl_sigmarek3} \\
{\Xi_j^{(1)}}' & = & e^{\mu_j^{(1)}} {k \choose j} 
  \Biggl\{ p\,\Xi_*^{(0)} \biggl( \sum_{l=0}^{k} e^{\mu_{l+1}^{(0)}}
    e^{-\mu_{l}^{(0)}} \Xi_l^{(0)}\biggr)^{p-1} \Biggr\}^j \times 
  \Biggl\{\biggl(\sum_{l=0}^{k}\Xi_l^{(0)} \biggr)^p \Biggr\}^{k-j} \,
  . \label{gl_sigmarek4}
\end{eqnarray}
\end{subequations}
The stroked partition functions belong to the rooted tree that results
from the iteration. We shall only explain the structure of
Eq.~\eqref{gl_sigmarek2} exemplarily. It describes the situation where
the root of the resulting tree is in the 0-lattice (see
Fig.~\ref{fig_ocemiter} left) and empty with $j$ occupied neighbouring
sites. The exponential prefactor accounts for the addition of a site
of this kind. The binomial factor counts the possibilities to
distribute the $j$ particles on the $k$ adjacent branches. The first
curly brackets refer to the situation of a branch containing a
particle. This particle can sit on the 1-lattice (first term in the
brackets, $\Xi_*^{(1)}$). The factor following $\Xi_*^{(1)}$ expresses
the possibilities for empty sites with a given number of occupied
neighbours on the other $p-1$ sites of the clique. Note that the factor
$e^{\mu_{l+1}^{(0)}} e^{-\mu_{l}^{(0)}}$ is due to the particle
present on the 1-site of the clique which increases the number of
occupied neighbours for the empty 0-sites by one. The second term in
the brackets stands for the $p-1$ possibilities of placing the
particle on a 0-site. The second curly brackets refer to an empty
clique. All possible combinations of empty sites with $l = 0, \ldots ,
k$ occupied neighbouring sites are included.

For the calculation of the partition functions we follow the same
lines as for Eqs.~\eqref{gl_rhoApproxIt}. We introduce intensive quantities with the
local fields   
\begin{equation}\label{gl_lokfeldsigma}
  x_j^{(0/1)} = \frac{\Xi_j^{(0/1)}}{\Xi_*^{(0/1)}}\: , \quad j=0,
  \ldots , k \, .
\end{equation}
By taking the ratio of Eq.~\eqref{gl_sigmarek2} and
Eq.~\eqref{gl_sigmarek1} as well as Eq.~\eqref{gl_sigmarek4} and
Eq.~\eqref{gl_sigmarek3} for $j=0, \ldots , k$ we obtain $2(k+1)$
equations for these local fields that still contain the $2(k+2)$
chemical potentials. These can be eliminated via the introduction of
physical partition functions $\tilde{\Xi}_j^{(0/1)}$, $j= 0, \ldots ,
k+1$ and $\tilde{\Xi}_*^{(0/1)}$ which replace the stroked quantities
in Eqs.~\eqref{gl_sigmarek} after substitution of $k \to k+1$ in the binomials
and exponents. By taking the ratios of Eqs.~\eqref{gl_sigmarek}
as before but for $j=0, \ldots , k+1$ and using that
$\frac{\tilde{\Xi}_j^{(0/1)}}{\tilde{\Xi}_*^{(0/1)}} =
\frac{\sigma_j^{(0/1)}}{\rho^{(0/1)}}$ we obtain $2(k+2)$ more
equations, so that we end up with the same number of equations and
unknown local fields and chemical potentials.

The solution of the corresponding equations for the
$\rho$-approximation was obtained by simple algebra which seems not
possible in the present case considering the complexity of the
equations. We can, however, show that the ansatz
\begin{subequations}
\label{gl_sigmajloes}
\renewcommand{\theequation}{\theparentequation \roman{equation}}
\begin{eqnarray}
  x_j^{(0/1)} & = & \alpha^{(0/1)} \, \frac{k+1-j}{k+1} \, \sigma_j^{(0/1)}
  \, , \label{gl_sigmajloes1} \\
  e^{\mu_{j+1}^{(0/1)}} e^{-\mu_{j}^{(0/1)}} x_j^{(0/1)} & = &
  \beta^{(0/1)} \, \frac{j+1}{k+1} \, \sigma_{j+1}^{(0/1)}  
  \label{gl_sigmajloes2}
\end{eqnarray}
\end{subequations}
with $j = 0, \ldots , k$ solves the equations under certain
conditions. One of coefficients $\alpha^{(0/1)}$ and $\beta^{(0/1)}$
can be chosen arbitrarily for each sublattice. These two extra degrees
of freedom arise from the consistency conditions that must be
imposed for the validity of Ansatz~\eqref{gl_sigmajloes} and which
read
\begin{equation}
 \rho^{(0)}  =  \frac{1}{p} \, \sum_{l=0}^{k} \frac{l+1}{k+1} \,
   \sigma_{l+1}^{(1)} \, , \qquad
 \rho^{(1)}  = \sum_{l=0}^{k} \frac{l+1}{k+1} \, \sigma_{l+1}^{(0)}
 - \frac{p-1}{p} \, \sum_{l=0}^{k} \frac{l+1}{k+1} \,
   \sigma_{l+1}^{(1)}  \label{gl_app1real}
\end{equation}
and which reduce the independent observables by one for each
sublattice. Eqs.~\eqref{gl_app1real} can be easily understood through
the observation that, e.g., empty sites on the 1-lattice with occupied
neighbouring sites contribute to the particle density on the 0-lattice
(first equation). Eqs.~\eqref{gl_app1real} are indeed important
conditions as they assure that a choice of $\sigma_j^{(0/1)}$ is
realisable within a microscopic configuration, and thus they do not
imply any restriction of Ansatz~\eqref{gl_sigmajloes} for physically
relevant situations.

As in Sec.~\ref{sec_rhoapprox} we now derive the differential equations for the
time evolution of the observables. Unknown probabilities are again
approximated by their pseudo-equilibrium values in the ensemble
characterised by $\Xi_G$. A complete list of the contributions due to the
different actions of the dynamics is given in Appendix~\ref{app_contrisigma}. We
shall only discuss the case of the jump from a 0-site to a 1-site
which qualitatively includes the calculations occurring in all other
cases.

\begin{figure}[tbp]
  \begin{center}
    \includegraphics[height = 7cm]{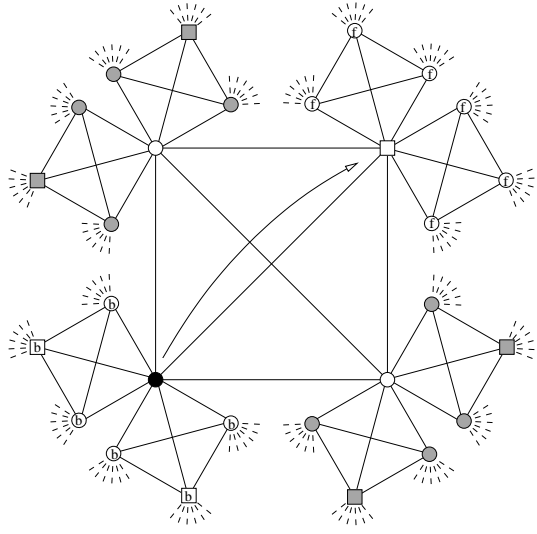}
    \caption{Sketch for the jump from a 0-site to a 1-site ($k=2, p=3$).}
\label{fig_01sprung}
  \end{center}
\end{figure}

The occurrence of a 0-1-jump requires that the algorithm selects a
0-site (1) which is occupied (2) and the decision to try a jump must
be made (3) in favour of an ending site which is in the
1-lattice (4). The probability is given by $(1) \cdot (2) \cdot (3)
\cdot(4) = \frac{p}{p+1} \cdot \rho^{(0)} \cdot q_s \cdot
\frac{1}{p}$. It remains to calculate the probability that in the
above situation the jump will be successful. To this purpose we
introduce the quantity $\Xi_e'^{(0/1)} = \sum_{l=0}^{k}
e^{\mu_{l+1}^{(0/1)}} e^{-\mu_l^{(0/1)}}\Xi_{l}^{(0/1)}$ which
corresponds to a rooted tree with an empty root but with the
modification that one extra occupied site is connected to the
root (empty sites connected to the root will not alter the
expression). Then the situation where (1) to (4) holds can be
described by
\begin{equation}
  \Xi_*^{(0)}\,(\Xi_e'^{(0)})^{p-1}\,\Xi_e'^{(1)} \, ,
\label{gl_cliquecombi}
\end{equation}
in terms of conditional partition functions. The situation is depicted in
Fig.~\ref{fig_01sprung} where for grey sites it is not known whether they are empty
or occupied. For a successful jump we must require all sites labelled
with f to be empty. This is a restriction of the above situation which
reads in terms of conditioned partition functions:  
\begin{equation}
  \Xi_*^{(0)}\,(\Xi_e'^{(0)})^{p-1}\, e^{\mu_{1}^{(1)}}
  e^{-\mu_0^{(1)}}\Xi_0^{(1)} \, .\label{gl_cliquecombifav}
\end{equation}

The ratio of Eqs. \eqref{gl_cliquecombifav} and \eqref{gl_cliquecombi}
gives the probability that the jump can actually be performed:
\begin{eqnarray}
 & & \frac{ \Xi_*^{(0)}\,(\Xi_e'^{(0)})^{p-1}\, e^{\mu_{1}^{(1)}}
    e^{-\mu_0^{(1)}}\Xi_0^{(1)} } {
    \Xi_*^{(0)}\,(\Xi_e'^{(0)})^{p-1}\,\Xi_e'^{(1)} } = \frac{
    e^{\mu_{1}^{(1)}} e^{-\mu_0^{(1)}}\Xi_0^{(1)}}{\sum_{l=0}^{k}
    e^{\mu_{l+1}^{(1)}} e^{-\mu_l^{(1)}} \Xi_{l}^{(1)}} \nonumber \\ &
    = & \frac{ e^{\mu_{1}^{(1)}}
    e^{-\mu_0^{(1)}}x_0^{(1)}}{\sum_{l=0}^{k} e^{\mu_{l+1}^{(1)}}
    e^{-\mu_l^{(1)}} x_{l}^{(1)}}
    \overset{\text{\eqref{gl_sigmajloes1}}}{=}
    \frac{\frac{1}{k+1}\sigma_1^{(1)}}{\sum_{l=0}^{k}
    \frac{l+1}{k+1}\,\sigma_{l+1}^{(1)}} \, .
\end{eqnarray}

We shall now discuss the contributions of the 0-1-jump to the changes in the
$\sigma_j^{(0/1)}$ which we divide in three classes. The first class
contains the {\em intra-clique contributions} which are all contributions coming from sites in the clique where the jump takes
place (i.e. the central clique in Fig.~\ref{fig_01sprung}). For the
$(p-1)$ empty 0-sites in this clique the number of occupied
neighbouring sites remains unaltered through the jump. A non-zero
intra-clique contribution is due to the occupied 0-site. After the
jump, this site is
empty with exactly one occupied neighbouring site which
means that the number of sites corresponding to $\sigma_1^{(0)}$ must
be increased by one. The second intra-clique contribution stems from
the ending site which has one occupied neighbour before the jump but
becomes an occupied site after the jump. This means we must decrease
the number of vacancies corresponding to $\sigma_1^{(1)}$ by one.

The second class of contributions will be termed as {\em backward
contributions}. They stem from the neighbouring sites which the particle
leaves behind during the jump (labelled with b in
Fig.~\ref{fig_01sprung}). The backward 0-sites may have $j = 0 ,
\ldots , k$ occupied neighbours in addition to the jumping
particle. The corresponding sites are not shown in
Fig.~\ref{fig_01sprung}, only the edges are drawn as dashed lines. We
will have a positive contribution to $\sigma_j^{(0)}$ from the sites
having $j$ occupied neighbours in addition to the jumping particle. The
same contribution with a negative sign must be attributed to
$\sigma_{j+1}^{(0)}$. In order to quantify the contribution, we need
to calculate the expected number of backward sites with $j$ extra
neighbours. For the calculation we define the partition function
$\Xi_{*,\nu_j}^{(0)}$ which refers to a 0-rooted tree with occupied root
having exactly $\nu_j$ neighbouring 0-sites with $j$ extra occupied neighbours. The
quantity is given by

\begin{equation}
  \Xi_{*,\nu_j}^{(0)}  =  {k(p-1) \choose \nu_j}
 \biggl(\sum_{l=0}^{k}e^{\mu_{l+1}^{(1)}}
 e^{-\mu_{l}^{(1)}}\Xi_l^{(1)}\biggr)^{k} \Bigl( e^{\mu_{j+1}^{(0)}}
    e^{-\mu_{j}^{(0)}}\Xi_j^{(0)} \Bigr)^{\nu_j}
 \biggl(\sum_{\substack{l=0\\l \neq j}}^{k}e^{\mu_{l+1}^{(0)}}
    e^{-\mu_{l}^{(0)}}\Xi_l^{(0)}  \biggr)^{k(p-1)-\nu_j} \, .
    \label{gl_01condnuj}
\end{equation}

The average number of backward 0-sites with $j$ additional occupied
neighbours is then given by
\begin{equation}
  \sum_{\nu_j = 0}^{k (p-1)} \nu_j \,
  \frac{\Xi_{*,\nu_j}^{(0)}}{{\Xi_*^{(0)}}'} = k(p-1)\,
   \frac{ \frac{j+1}{k+1} \, \sigma_{j+1}^{(0)}}
   {\sum_{l=0}^{k}\frac{l+1}{k+1} \, \sigma_{l+1}^{(0)}} \, . 
\end{equation}

Here ${\Xi_*^{(0)}}'$ serves as a normalisation as it corresponds to
the unconstrained scenario. The r.h.s. is obtained by substitution of
${\Xi_*^{(0)}}'$ and $\Xi_{*,\nu_j}^{(0)}$ according to
Eqs.~\eqref{gl_sigmarek1} and \eqref{gl_01condnuj} and performing the
summation. With this result the backward contribution of the 0-sites
can be easily expressed (see Appendix~\ref{app_contrisigma}). The calculation of the
backward contribution of the 1-sites follows very similar lines
yielding an average number of backward 1-sites with $j$ additional
occupied neighbouring sites which is
\begin{equation}
 k\, \frac{ \frac{j+1}{k+1} \, \sigma_{j+1}^{(1)}}
   {\sum_{l=0}^{k}\frac{l+1}{k+1} \, \sigma_{l+1}^{(1)}} \, .
\end{equation}

A third class contains the {\em forward contributions} which come from
the sites labelled with f in Fig.~\ref{fig_01sprung}. The f-sites
having exactly $j-1$ occupied neighbouring sites will yield a positive
contribution to $\sigma_j^{(0)}$ and a negative contribution comes
from the f-sites with exactly $j$ occupied neighbouring sites. The
average number of f-sites with exactly $j$ occupied neighbouring sites
can be calculated with the help of the corresponding partition function
\begin{equation}
  \Xi_{0, \nu_j}^{(1)} =  e^{\mu_0^{(1)}} {kp \choose \nu_j} \Bigl(
  \Xi_j^{(0)} \Bigr)^{\nu_j} 
  \biggl(\sum_{\substack{l=0\\l \neq j}}^{k}\Xi_l^{(0)}
  \biggr)^{kp-\nu_j} \, .
\end{equation}
One obtains the average number as
\begin{equation}
  \sum_{\nu_j=0}^{kp} \nu_j \,
  \frac{\Xi_{0,\nu_j}^{(1)}}{{\Xi_0^{(1)}}'} = kp \,
  \frac{\frac{k+1-j}{k+1}\,\sigma_j^{(0)}}
  {\sum_{l=0}^{k}\frac{k+1-l}{k+1}\,\sigma_l^{(0)}} \, .
\end{equation}

With similar arguments all the contributions can be derived (for the
results see Appendix~\ref{app_contrisigma}) and one ends up with a system of $2(k+2)$
differential equations which is of the form 
\begin{equation}\label{gl_sigmajdiffgl}
\dot{\sigma}_j^{(\iota)}  =  \dot{\sigma}_j^{(\iota)}\bigl(\sigma_0^{(0)}, \ldots ,
    \sigma_{k+1}^{(0)}; \sigma_0^{(1)}, \ldots ,
    \sigma_{k+1}^{(1)}\bigr)\, ,
\end{equation}
with $j = 0, \ldots , k+1$ and $\iota = 0 , 1$.

For given initial conditions  $\rho^{(0)}(t=0)=\rho^{(0)}$ and
$\rho^{(1)}(t=0)=\rho^{(1)}$ we first calculate the initial values
$\sigma_j^{(0/1)}(t=0)$ which can be performed following similar lines
to the calculation of $(1-\rho^{(0/1)})\mathcal{P}_1^{(0/1)}$
(see Sec.~\ref{sec_rhoapprox}, Eq.~\eqref{gl_P10rechnung}).
$(1-\rho^{(0/1)})\mathcal{P}_1^{(0/1)}$ is the probability for an
empty site without occupied neighbouring sites which is obviously
equivalent to $\sigma_0^{(0/1)}$. A more general calculation yields,
for $j = 0, \ldots , k+1$,
\begin{equation}\label{gl_sigmajanfbed}
  \sigma_j^{(0/1)}(t=0) \, = \,\bigl(1-\rho^{(0/1)}\bigr){k+1 \choose
    j}   
  \left(1-\frac{1-\rho^{(1)}-p\,\rho^{(0)}}{1-\rho^{(0/1)}}\right)^{j}
  \left(\frac{1-\rho^{(1)}-p\,\rho^{(0)}}{1-\rho^{(0/1)}}\right)^{k+1-j}
  \ .
\end{equation}
With these initial values, the $\sigma_j$-approximation is obtained by
integrating Eqs.~\eqref{gl_sigmajdiffgl} numerically. We add that the
conditions \eqref{gl_app1real} are fulfilled by the initial values and
preserved by Eqs.~\eqref{gl_sigmajdiffgl}.

Eqs.~\eqref{gl_sigmajdiffgl} can be evaluated asymptotically as this
was done analytically for the $\rho$-approximation in
Sec.~\ref{sec:rhoasymp}. The Jacobian matrix $A$ of the system is a
$2(k+2) \times 2(k+2)$-matrix:
\begin{equation}
  A = \left(
       \begin{array}{cc}
         A_{00} & A_{01} \\
         A_{10} & A_{11}
       \end{array}
 \right) \, ,
\qquad
\text{where}
\quad
A_{mn} = \left (
  \begin{array}{ccc}
     \frac{\partial \dot{\sigma}_0^{(m)}}{\partial \sigma_0^{(n)}} 
   & \ldots & \frac{\partial \dot{\sigma}_0^{(m)}}{\partial
     \sigma_{k+1}^{(n)}} \\ 
     \vdots & \vdots & \vdots \\
    \frac{\partial \dot{\sigma}_{k+1}^{(m)}}{\partial \sigma_0^{(n)}} 
   & \ldots & \frac{\partial \dot{\sigma}_{k+1}^{(m)}}{\partial
     \sigma_{k+1}^{(n)}}
  \end{array}
\right ) \, .
\end{equation}

The matrix can be calculated with the help of a computer algebra
system as well as its evaluation at a stationary points of
Eqs.~\eqref{gl_sigmajdiffgl} can be done. We find that two eigenvalues
always vanish which is due to the conditions \eqref{gl_app1real} that
reduce the effective number of observables by two. If the other
eigenvalues $\{\lambda_i\}$ are all negative, the stationary point is
stable and the inverse relaxation time is given by $\tau^{-1} =
-\max\{\lambda_i\}$.

In the next section we show results for relaxation times and for the
time-flow of the observables according to the $\rho$-approximation and
$\sigma_j$-approximation.

\subsection{Comparison with MC simulations}
\label{sec:MCcrystal}

As the exact dynamics is not accessible so far, there is no intrinsic way
to check the quality of the results obtained according to the PAS. We
therefore performed MC~simulations implementing the dynamics defined
in Sec.~\ref{sec_defdyn} on large graphs,
details of the generation of the graphs were already given in
\cite{HaWe}.

\subsubsection*{Equilibration time}

As a first example, we look at the relaxation time for a system with
$k=2$ and $p=1$, i.e. on an ordinary Bethe lattice with coordination
number three. On this lattice, due to $p=1$, the crystallisation
transition is of second order. The simulations where performed for an
average annihilation rate $q_0=0.2$ and a mobility of $q_s = 0.8$. The
parameter $q_1$ was used to tune the chemical potential $\mu$
according to the condition (\ref{eq:detailled}) of detailed balance.

Let us shortly discuss the results of the analytical
$\rho$-approximation for these parameters. For very diluted systems
($q_1\to 0, \ \mu\to-\infty$), the relaxation time goes to
$1/q_0$. For increasing $\mu$ the relaxation time decreases until it
reaches a local minimum at $\mu_{eq}\simeq -0.445$, the point where
the slowest process changes from being the decay of global density
fluctuations to the decay of differences between the sublattice
densities. The equilibration time starts to grow until it diverges as
$|\mu-\mu_{cr}|^{-1}$ if we approach the critical point $\mu_{cr}\simeq
1.383$. It restarts to decrease and at the end saturates at
$1/\tau_{eq}(\mu\to\infty) = q_0 + q_s/p$ which equals one for the
above choice of parameters.

The $\sigma_j$-approximation leads to qualitatively similar, but
quantitatively slightly different results: It starts at the same value
of the equilibration time for $\mu\to-\infty$, but reaches the local
minimum already at $\mu_{eq}\simeq -0.556$. The behaviour close to the
phase transition is extremely similar, with a slightly different
prefactor to the dominant critical divergence. At large chemical
potentials, we have $1/\tau_{eq}(\mu\to\infty) = \frac 34 q_0 +
q_s=0.95$, i.e. the equilibration time in the vicinity of the closest
packing is slightly larger than predicted by the simpler
approximation.

\begin{figure}
  \begin{center}
    \begin{tabular}{cc}
      \includegraphics[height = 5.5cm]{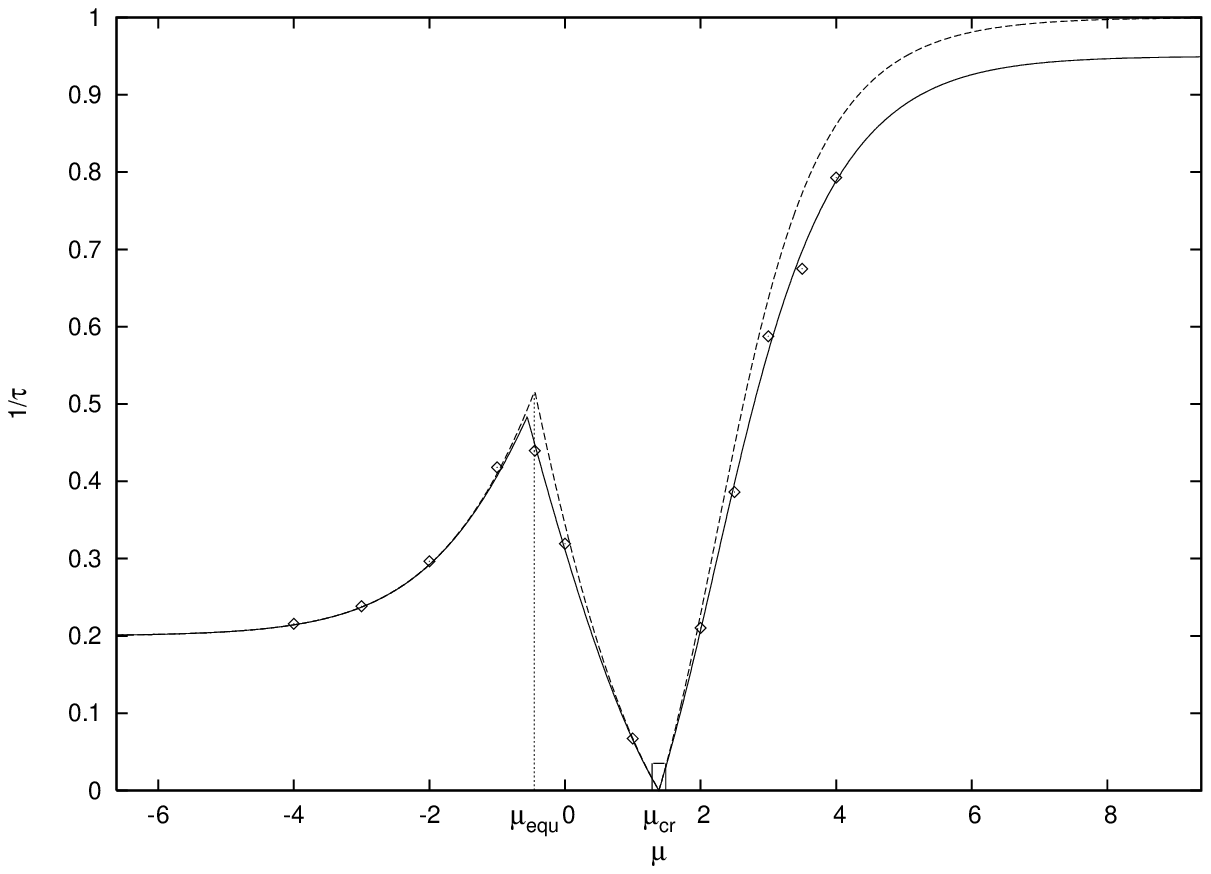} &
      \includegraphics[height = 5.5cm]{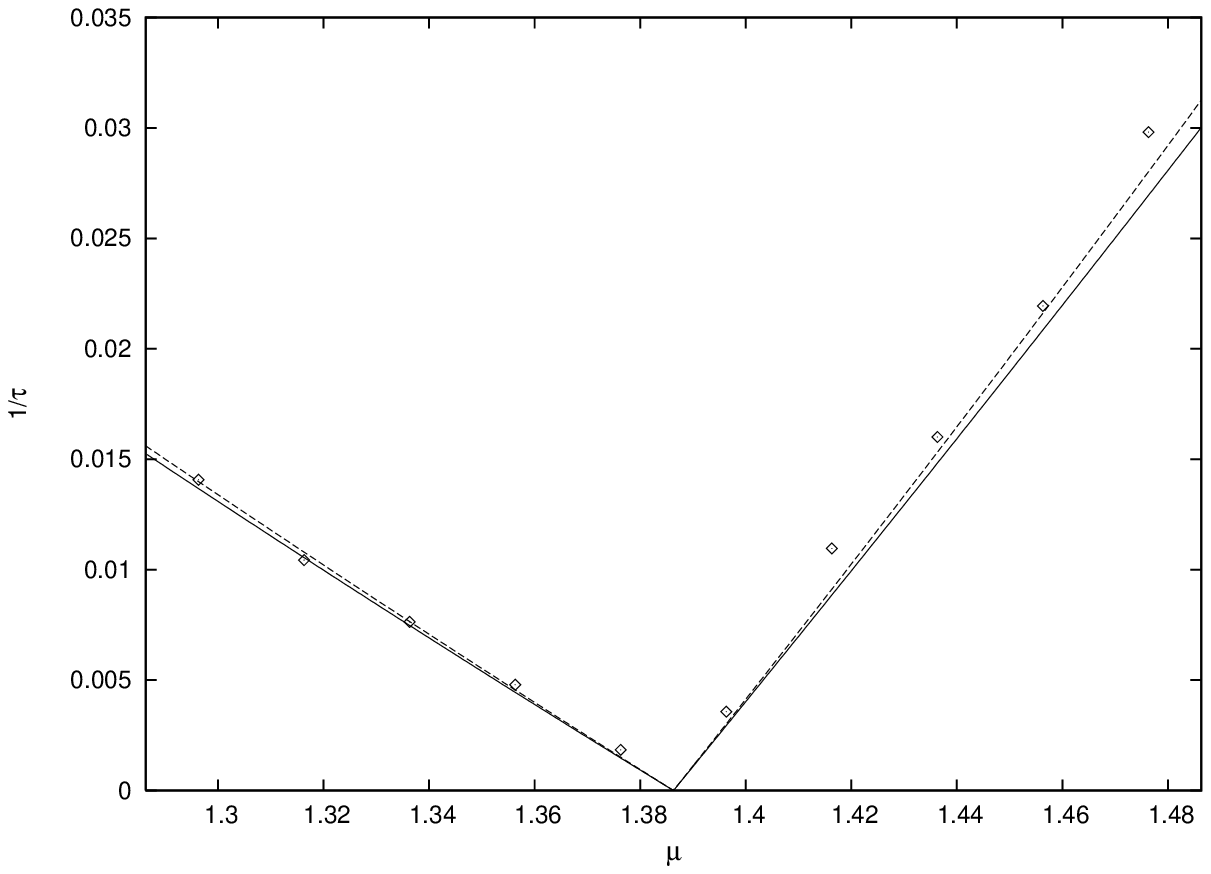}
    \end{tabular}
   \caption{Inverse relaxation time for $q_0 = 0.2$ and $q_s = 0.8$.
      The dashed line shows the results of the $\rho$-approximation,
      the full line those of the $\sigma_j$-approximation. MC data is
      represented by the symbols.  The right figure enlarges the
      critical region to show the linear vanishing of $1/\tau$. The
      slopes on both sides of the transition are related by a
      factor two.}
   \label{fig_relaxall}
  \end{center}
\end{figure}

For comparison, we have plotted these results together with numerical
data in Fig.~\ref{fig_relaxall}. The numerical simulations were
performed on a large bipartite lattice of $N=5\cdot 10^6$ sites, as an
initial condition we have chosen one densest packing, the data were
averaged over 10 independent runs. For extracting the equilibration
time, we have plotted the logarithm of the difference
$\rho^{(1)}(t)-\rho^{(1)}_{stat}$ from the analytically known
equilibrium value, as a function of the MC time. The function was
linearised in a suitable region. The problem hereby is to identify
this region: For short times, the system is still far from its
asymptotic behaviour, for large time the dynamics is dominated by
fluctuations. The estimated error of this procedure grows up to
10\% for the values close to the transition point.

Fig.~\ref{fig_relaxall} demonstrates that both approximations show a
very good qualitative agreement with the numerical findings. The more
detailed $\sigma_j$-approximation gives, however, the better
quantitative values which, within the estimated error, almost coincide with
the MC results.

\subsubsection*{Equilibration of the sublattice densities}

As a further comparison we have investigated the relaxation of the
sublattice densities for a system with $k=3$ and $p=2$.  We have
performed the numerical experiments on a graph of $N = 4.5\cdot 10^6$
vertices, and have chosen the dynamical parameters as $q_1=1$,
$q_0=e^{-2}$ and $q_s = 1 - q_0$. These values correspond to a
chemical potential $\mu = 2$ which is situated in the interval where
the liquid as well as the normal crystalline phase are locally stable.

\begin{figure}[htbp]
  \begin{center}
    \includegraphics[height = 8cm]{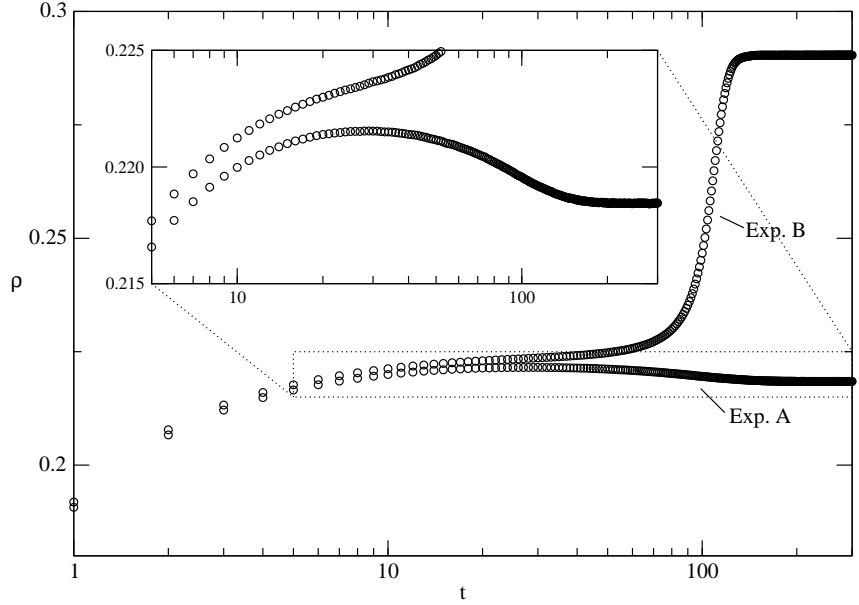}
    \caption{Global density for two systems ($k=3$, $p=2$) with
    initial conditions $\rho^{(0)}(t=0)=0$ and $\rho^{(1)}(t=0)=0.145$
    (experiment A) resp. $\rho^{(1)}(t=0)=0.160$ (experiment B).
    The asymptotically reached densities are those of the two
    metastable states of the system, the smaller one corresponding to
    the liquid, the larger one to the crystalline phase. Numerical
    data was averaged over 100 runs. }
\label{fig_totdenlicr}
  \end{center}
\end{figure}

We have chosen two slightly different initial conditions, in both
cases the $0$-sublattice was empty. The $1$-sublattice was first
initialised with a partial density $\rho^{(1)}(t=0)=0.145$ (experiment
A), in the second case with $\rho^{(1)}(t=0)=0.160$ (experiment
B). After a similar evolution in the initial time interval, the
systems approach the two different locally stable solutions of the
model. Whereas the less dense one converges to the liquid solution,
the denser one crystallises. The local stability of the liquid phase
is demonstrated impressively in case A: After an initial increase, the
global density falls again toward the liquid value.

\begin{figure}[htbp]
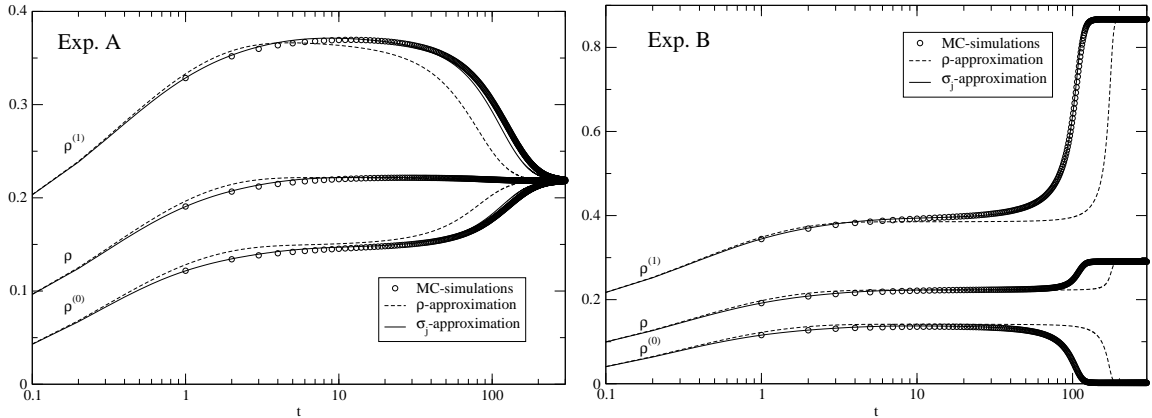

  \begin{center}
    \begin{tabular}{cc}
      \includegraphics[height = 5.5cm]{liquid.eps} &
      \includegraphics[height = 5.5cm]{crystl.eps}
    \end{tabular}
    \caption{Global and sublattice densities for the two
    approximations and the MC simulation. For the corresponding
    initial conditions see caption of Fig.~\ref{fig_totdenlicr}.}
\label{fig_pardenli}
  \end{center}
\end{figure}

In Fig.~\ref{fig_pardenli}, the numerical results for the global and
both sublattice densities are compared to the $\rho$- and the
$\sigma_j$-approximation. As in the case of the equilibration times,
both approximations reflect correctly the qualitative behaviour. The
simpler $\rho$-approximation, however, shows a rather pronounced
difference in the onset of the final relaxation toward the stationary
point. This is almost completely cured by the more involved
$\sigma_j$-approximation, which shows a quantitatively impressing
agreement with the MC data over the whole time interval.

\section{Dynamics at the glass-transition}
\label{sec:glass}

The aim of this section is to detect also the glass-transition in the
context of PAS, i.e.~through its impact on time-dependent, dynamical
quantities. Apparently, the observables considered so far (essentially
particle densities) are not suitable to probe the fundamental
difference between the liquid and amorphous state. This is clear as,
e.g., the difference between the liquid solution densities and the
1RSB solution densities is hardly significant at least for an
amorphous system in proximity to the glass-transition. It seems more
promising to focus on the long lifetime of arrested configurations in
the glassy state which obviously strongly affects the two-time
autocorrelation function
\begin{equation}\label{gl_zweizeitauto}
  C(t_1, t_2) = \frac{1}{N}\sum_{i = 1}^{N} \delta_{n_i(t_1),
    n_i(t_2)} \quad , \qquad t_1 \le t_2 \, ,
\end{equation}
where  $n_i(t)\in\{0, 1\}$ is the occupation number of site $i$ at
time $t$. In the case of a liquid system we expect $C$ to decay quickly with
increasing $t_2-t_1$ to the value $C_{\text{li}} = \rho^2+(1-\rho)^2$
that corresponds to uncorrelated configurations.

For a more symmetric approach we slightly modify the definition of $C$
in Eq.~\eqref{gl_zweizeitauto} from a two-time autocorrelation to
inter-system correlation of two copies of the original model. We
introduce
\begin{equation}
  C(t) = \frac{1}{N}\sum_{i=1}^N \delta_{n_i^{I}(t), n_i^{I\!I}(t)} \, ,
\end{equation}
where $n_i^{I}(t)$ and $n_i^{I\!I}(t)$ are the occupation numbers at
time $t$ of two systems indexed $I$ and $I\!I$ that are defined on the
same graph. Apart from their topological structure the systems can be
different, in particular we will consider the case where both systems
are characterised by different dynamical rates $q_0,q_1$ and $q_s$.
We recover Def.~\eqref{gl_zweizeitauto} for a fixed first system
with $n_i^{I}(t)\doteq n_i^{I\!I}(t_1)$, and by setting $t\doteq
t_2$. The correlation function $C(t)$ is the quantity we shall
approximate in the following.

\subsection{The correlated-$\rho$-approximation}

The use of $C(t)$ as a single observable for PAS seems insufficient
for the treatment of correlated systems. We therefore transport the
simplest description of a liquid system which is given by the particle
density $\rho$ to the level of two correlated systems $I$ and $I\!I$
which are defined on isomorphic lattices (i.e. every site has a
partner site in the other system). This can be done by counting the
number of isomorphic pairs of sites that are (a) both empty, (b)
occupied in system $I$ but empty in system $I\!I$, (c) empty in system
$I$ and occupied in system $I\!I$ and (d) occupied in both
systems. The corresponding densities w.r.t. to the number of
isomorphic pairs (i.e. $N$) are denoted by $\rho_{ee}$, $\rho_{e*}$,
$\rho_{*e}$ and $\rho_{**}$. They constitute the minimal set of
observables we shall apply PAS to. The correlation function $C(t)$ is
then obtained via
\begin{equation}\label{gl_zshgcorrrho0011}
C(t) = \rho_{ee}(t) + \rho_{**}(t) \, .
\end{equation}

In a first step we introduce the generalised pseudo-equilibrium by
defining three conjugate chemical potentials $\mu_{e*}$, $\mu_{*e}$
and $\mu_{**}$. A chemical potential $\mu_{ee}$ is not required as
$\rho_{ee} = 1-\rho_{e*}-\rho_{*e}-\rho_{**}$ is not an independent
observable. Analogously to the previous section, we introduce
conditional partition functions $\Xi_{ee}$, $\Xi_{e*}$, $\Xi_{*e}$ and
$\Xi_{**}$ which refer to isomorphic rooted trees with an isomorphic
pair on the root that is given by the index. The iteration of rooted
trees can be performed as usual with the difference that we must
iterate simultaneously in the two systems now. In terms of conditional
partition functions this reads
  \begin{subequations}
\label{gl_rekcorr}
\renewcommand{\theequation}{\theparentequation \roman{equation}}
\begin{eqnarray}
  {\Xi_{ee}}' & = & \Bigl(\Xi_{ee}^p + p\,
    \Xi_{e*}\Xi_{ee}^{p-1} + p\,
    \Xi_{*e}\Xi_{ee}^{p-1}+ p\,\Xi_{**}\Xi_{ee}^{p-1} +
    p(p-1)\Xi_{e*}\Xi_{*e}\Xi_{ee}^{p-2}  \Bigr)^k \label{gl_rekcorrbsp}\\
  {\Xi_{e*}}' & = &
  e^{\mu_{e*}}\Bigl(\Xi_{ee}^p+p\,\Xi_{*e}\Xi_{ee}^{p-1} \Bigr)^k
  \label{gl_rekcorr2}\\
  {\Xi_{*e}}' & = &
  e^{\mu_{*e}}\Bigl(\Xi_{ee}^p+p\,\Xi_{e*}\Xi_{ee}^{p-1} \Bigr)^k
  \label{gl_rekcorr3}\\
  {\Xi_{**}}' & = & e^{\mu_{**}}\Bigl(\Xi_{ee}^p\Bigr)^k \, .
\end{eqnarray}
\end{subequations}
We only explain Eq.~\ref{gl_rekcorrbsp}, i.e. the case where the root
of the resulting tree is empty in the two systems. This implies the
following possibilities for each pair of adjacent cliques (content of
large brackets): (1) both cliques are empty; (2) clique $I$ is empty
and clique $I\!I$ carries a particle; (3) as (2) with $I
\leftrightarrow I\!I$; both cliques carry a particle which are on
isomorphic sites (4) or on sites that are not isomorphic (5). The
contributions must by weighted according to the number of
realisations:  $(1) + p\cdot(2)+p\cdot(3)+p\cdot(4)+p(p-1)\cdot(5)$.

For the calculation we introduce intensive quantities (local fields):
\begin{equation}
  x \doteq \frac{\Xi_{e*}}{\Xi_{ee}} \, , \quad y \doteq
  \frac{\Xi_{*e}}{\Xi_{ee}} \, , \quad z \doteq
  \frac{\Xi_{**}}{\Xi_{ee}} \, .
\end{equation}
We can obtain equations for the local fields by taking the
corresponding ratios of Eqs.~\eqref{gl_rekcorr}. These read
\begin{subequations}
\label{gl_corrlokfeld}
\renewcommand{\theequation}{\theparentequation \roman{equation}}
\begin{eqnarray}
 x & = & \frac{e^{\mu_{e*}}(1+py)^k}{\left(1+px+py+pz+p(p-1)xy\right)^k}\\
 y & = & \frac{e^{\mu_{*e}}(1+px)^k}{\left(1+px+py+pz+p(p-1)xy\right)^k}\\
 z & = & \frac{e^{\mu_{**}}}{\left(1+px+py+pz+p(p-1)xy\right)^k} \, .
\end{eqnarray}
\end{subequations}

They still contain the unknown chemical potentials which we eliminate
as described in previous calculations via the introduction of physical
partition functions (i.e. the substitution $k \rightarrow k+1$). This
yields the equations
\begin{subequations}
\label{gl_corrphys}
\renewcommand{\theequation}{\theparentequation \roman{equation}}
\begin{eqnarray}
  \frac{\rho_{e*}}{\rho_{ee}} & = & 
  \frac{e^{\mu_{e*}}(1+py)^{k+1}}{\left(1+px+py+pz+p(p-1)xy\right)^{k+1}}\\
  \frac{\rho_{*e}}{\rho_{ee}} & = & 
  \frac{e^{\mu_{*e}}(1+px)^{k+1}}{\left(1+px+py+pz+p(p-1)xy\right)^{k+1}}\\
  \frac{\rho_{**}}{\rho_{ee}} & = & 
  \frac{e^{\mu_{**}}}{\left(1+px+py+pz+p(p-1)xy\right)^{k+1}} \, .
\end{eqnarray}
\end{subequations}
Substituting the chemical potentials in Eqs.~\eqref{gl_corrlokfeld}
according to Eqs.~\eqref{gl_corrphys} we find
\begin{subequations}
\renewcommand{\theequation}{\theparentequation \roman{equation}}
\begin{eqnarray}
  \frac{\rho_{e*}}{\rho_{ee}} & = & \frac{x(1+py)}{1+px+py+pz+p(p-1)xy}\label{gl_xyz1}\\
  \frac{\rho_{*e}}{\rho_{ee}} & = & \frac{y(1+px)}{1+px+py+pz+p(p-1)xy}\label{gl_xyz2}\\
  \frac{\rho_{**}}{\rho_{ee}} & = & \frac{z}{1+px+py+pz+p(p-1)xy} \,
  .\label{gl_xyz3}
\end{eqnarray}
\end{subequations}

This results in a quadratic equation for $x$ when we
substitute $y$ and $z$ in Eq.~\eqref{gl_xyz1} according to
Eqs.~\eqref{gl_xyz2} and \eqref{gl_xyz3}. Only one of the two
solutions for $x$ is consistent with the constraints that apply to the
observables (positivity and restriction of the particle density by
$\frac{1}{p+1}$). We finally end up with a unique solution for $x$, $y$
and $z$ which is
\begin{subequations}
\label{gl_lsgxyz}
\renewcommand{\theequation}{\theparentequation \roman{equation}}
\begin{eqnarray}\label{gl_lsgxyzx}
 x & = & \frac{\sqrt{1+4p\alpha\beta\,}+2p\alpha-1}{2p(1+\beta-p\alpha)}\\
 y & = &
 \frac{\sqrt{1+4p\alpha\beta\,}+2p\beta-1}{2p(1+\alpha-p\beta)}\\
 z & = & \gamma (1+px+py+p(p-1)xy) \, ,
\end{eqnarray}
\end{subequations}
where
\begin{equation}
  \alpha = \frac{\rho_{e*}}{\rho_{ee}-p\rho_{**}}\, , \quad
  \beta = \frac{\rho_{*e}}{\rho_{ee}-p\rho_{**}} \, \quad\text{and}\quad
  \gamma = \frac{\rho_{**}}{\rho_{ee}-p\rho_{**}}\, .
\end{equation}

For the derivation of the differential equations for the observables
we only discuss four typical cases. The described actions always refer
to system $I$:
\begin{itemize}
\item
{\em Insertion on an $ee$-site:} The partition function for an
$ee$-site inside the graph is obtained by substituting $k \rightarrow
k+1$ in Eq.~\eqref{gl_rekcorrbsp}. For a successful insertion we must
require that all neighbouring sites of the $ee$-site in system $I$ are
empty which corresponds to the partition function $(\Xi_{ee}^p + p\,
\Xi_{e*}\Xi_{ee}^{p-1})^{k+1}$. The corresponding probability is
obtained by dividing by the partition function for an unconstrained
$ee$-site:
\begin{equation}
\frac{(\Xi_{ee}^p + p\,\Xi_{e*}\Xi_{ee}^{p-1})^{k+1}}{(\Xi_{ee}^p + p\,
    \Xi_{e*}\Xi_{ee}^{p-1} + p\,
    \Xi_{*e}\Xi_{ee}^{p-1}+ p\,\Xi_{**}\Xi_{ee}^{p-1} +
    p(p-1)\Xi_{e*}\Xi_{*e}\Xi_{ee}^{p-2})^{k+1}}
  = \left( \frac{1 + p x}{1 + px + py + pz +
    p(p-1)xy} \right )^{k+1} \, .\nonumber
\end{equation}
\item
{\em Insertion on an $e*$-site:} The partition function for an $e*$-site
follows from Eq.~\eqref{gl_rekcorr2} with $k \rightarrow k+1$. The
restriction that all neighbouring sites are empty has the partition
function $e^{\mu_{e*}}(\Xi_{ee}^p)^{k+1}$ leading to a probability for
a successful insertion which is
\begin{equation}
  \frac{e^{\mu_{e*}}(\Xi_{ee}^p)^{k+1}}{e^{\mu_{e*}}(\Xi_{ee}^p+p\,
    \Xi_{*e}\Xi_{ee}^{p-1})^{k+1}} = 
  \left(\frac{1}{1+py}\right)^{k+1} \, . \nonumber
\end{equation}
\item
{\em Jump from a $*e$-site to an $e*$-site:} The partition function for
an unconstrained $*e$-site follows from Eq.~\eqref{gl_rekcorr3} with
$k \rightarrow k+1$. We ask for the probability that the ending site
is $e*$ (and not $ee$). The corresponding partition function is
$(\Xi_{e*}\Xi_{ee}^{p-1})\,{\Xi_{*e}}'$ leading to the probability
\begin{equation}
  \frac{(\Xi_{e*}\Xi_{ee}^{p-1})\,{\Xi_{*e}}'}{(\Xi_{ee}^p+p\,
\Xi_{e*}\Xi_{ee}^{p-1})\,{\Xi_{*e}}'} = \frac{x}{1+px} \nonumber
\end{equation}
for an ending site which is $e*$. For a successful jump the ending
site in system $I$ must not have any other occupied neighbouring
sites. The corresponding partition function is
$e^{\mu_{e*}}(\Xi_{*e}\Xi_{ee}^{p-1})(\Xi_{ee}^p)^{k}$ where the
ending $e*$-site is regarded as the central site (exponential factor)
and the first brackets stand for the clique where the jump is
performed. The unconstrained case corresponds to
$e^{\mu_{e*}}(\Xi_{*e}\Xi_{ee}^{p-1})(\Xi_{ee}^p+p\,\Xi_{*e}
\Xi_{ee}^{p-1})^{k}$ which leads to the following probability for a
successful jump:
\begin{equation}
  \frac{e^{\mu_{e*}}(\Xi_{*e}\Xi_{ee}^{p-1})(\Xi_{ee}^p)^{k}}{
    e^{\mu_{e*}}(\Xi_{*e}\Xi_{ee}^{p-1})
(\Xi_{ee}^p+p\,\Xi_{*e}\Xi_{ee}^{p-1})^{k}} =
\left(\frac{1}{1+py}\right)^k \, . \nonumber
\end{equation}
\item
{\em Jump from a $**$-site to an $ee$-site:} In this case the ending
site is always $ee$ as the isomorphic partner of the starting site is
also occupied. The general partition function describing the situation
is $(\Xi_{**}\Xi_{ee}^{p-1})\,{\Xi_{ee}}'$ where the ending site
chosen as the central site and the first brackets correspond again to
the clique where the jump is performed. The case where the ending site
has no other occupied neighbouring sites corresponds to
$(\Xi_{**}\Xi_{ee}^{p-1})\,(\Xi_{ee}^p + \,\Xi_{e*}\Xi_{ee}^{p-1})^k$
leading to a probability for a successful jump which is
\begin{equation}
  \frac{(\Xi_{**}\Xi_{ee}^{p-1})\,(\Xi_{ee}^p +
p\,\Xi_{e*}\Xi_{ee}^{p-1})^k}{(\Xi_{**}\Xi_{ee}^{p-1})\,{\Xi_{ee}}'} =
\left(\frac{1+px}{1+px+py+pz+p(p-1)xy} \right)^k \, .\nonumber
\end{equation}
\end{itemize}
The corresponding expressions for system $I\!I$ can be obtained by
exchanging $x$ and $y$.

For the construction of the rate-equations for the observables we must
check how the different actions of the dynamics contribute to the change in the
observables. With the probabilities for the actions which are
calculated as shown above, we finally obtain
{\allowdisplaybreaks
\begin{subequations}
\label{gl_corrapprox}
\renewcommand{\theequation}{\theparentequation \roman{equation}}
\begin{eqnarray} \label{gl_corrapprox00}
  \dot{\rho}_{ee} &=  & + \rho_{*e}\,q_0^I -\rho_{ee}\,q_1^I\,
  \left(\frac{1+px}{1+px+py+pz+p(p-1)xy}
  \right)^{k+1}
  \nonumber \\*
   & & +  \rho_{e*}\,q_0^{I\!I}
       -\rho_{ee}\,q_1^{I\!I}\,\left(\frac{1+py}{1+px+py+pz+p(p-1)xy}
  \right)^{k+1} \nonumber \\*
   & & +
   \rho_{*e}\,q_s^I\,\frac{x}{1+px}\,\left(\frac{1}{1+py}\right)^k
   - \rho_{**}\,q_s^I\,\left(\frac{1+px}{1+px+py+pz+p(p-1)xy}\right)^k
   \nonumber \\*
   & & +
   \rho_{e*}\,q_s^{I\!I}\,\frac{y}{1+py}\,\left(\frac{1}{1+px}\right)^k
   -
   \rho_{**}\,q_s^{I\!I}\,\left(\frac{1+py}{1+px+py+pz+p(p-1)xy}\right)^k
 \nonumber \\*
\\
  \dot{\rho}_{e*} & = &  + \rho_{**}\,q_0^I
  -\rho_{e*}\,q_1^I\,\left(\frac{1}{1+py}
  \right)^{k+1} 
  \nonumber \\*
   & & - \rho_{e*}\,q_0^{I\!I}
       +\rho_{ee}\,q_1^{I\!I}\,\left(\frac{1+py}{1+px+py+pz+p(p-1)xy}
  \right)^{k+1} \nonumber \\*
   & &  
   -\rho_{*e}\,q_s^I\,\frac{x}{1+px}\,\left(\frac{1}{1+py}\right)^k
   + \rho_{**}\,q_s^I\,\left(\frac{1+px}{1+px+py+pz+p(p-1)xy}\right)^k  
   \nonumber \\*
   & &    
  -\rho_{e*}\,q_s^{I\!I}\,\frac{y}{1+py}\,\left(\frac{1}{1+px}\right)^k 
  + \rho_{**}\,q_s^{I\!I}\,\left(\frac{1+py}{1+px+py+pz+p(p-1)xy}\right)^k
 \nonumber \\* \\
  \dot{\rho}_{*e} & = &  - \rho_{*e}\,q_0^I
  +\rho_{ee}\,q_1^I\,\left(\frac{1+px}{1+px+py+pz+p(p-1)xy}
  \right)^{k+1}
  \nonumber \\*
   & & + \rho_{**}\,q_0^{I\!I}
  -\rho_{*e}\,q_1^{I\!I}\,\left(\frac{1}{1+px}
  \right)^{k+1} 
  \nonumber  \\*
   & &    
  -\rho_{*e}\,q_s^I\,\frac{x}{1+px}\,\left(\frac{1}{1+py}\right)^k 
  + \rho_{**}\,q_s^I\,\left(\frac{1+px}{1+px+py+pz+p(p-1)xy}\right)^k
 \nonumber \\*
   & &    
  -\rho_{e*}\,q_s^{I\!I}\,\frac{y}{1+py}\,\left(\frac{1}{1+px}\right)^k 
  + \rho_{**}\,q_s^{I\!I}\,\left(\frac{1+py}{1+px+py+pz+p(p-1)xy}\right)^k
 \nonumber \\* \\ \label{gl_corrapprox11}
  \dot{\rho}_{**} & = & - \rho_{**}\,q_0^I
  +\rho_{e*}\,q_1^I\,\left(\frac{1}{1+py}
  \right)^{k+1} 
  \nonumber \\*
   & & - \rho_{**}\,q_0^{I\!I}
  +\rho_{*e}\,q_1^{I\!I}\,\left(\frac{1}{1+px}
  \right)^{k+1} 
  \nonumber  \\*
   & & +
   \rho_{*e}\,q_s^I\,\frac{x}{1+px}\,\left(\frac{1}{1+py}\right)^k
   - \rho_{**}\,q_s^I\,\left(\frac{1+px}{1+px+py+pz+p(p-1)xy}\right)^k
   \nonumber \\*
   & & +
   \rho_{e*}\,q_s^{I\!I}\,\frac{y}{1+py}\,\left(\frac{1}{1+px}\right)^k
   -
   \rho_{**}\,q_s^{I\!I}\,\left(\frac{1+py}{1+px+py+pz+p(p-1)xy}\right)^k
 \nonumber \\*
\end{eqnarray}
\end{subequations}
}

Here $q_0^{I/I\!I}$ $q_1^{I/I\!I}$ and $q_s^{I/I\!I}$ denote the
probabilities for particle annihilation, creation and jumps in the
respective system.

Applying normalisation $\rho_{ee}+\rho_{e*}+\rho_{*e}+\rho_{**}=1$ and
the relations $\rho_{*e}+\rho_{**}=\rho^{I}$,
$\rho_{e*}+\rho_{**}=\rho^{I\!I}$, where $\rho^{I}$ and $\rho^{I\!I}$
are the particle densities in the respective system, as well as
Eq.~\eqref{gl_zshgcorrrho0011} we find that at any time $t$:
\begin{equation}\label{gl_zshgCrho}
  \begin{array}{ll}
  \rho_{ee}(t) = \frac{1}{2}\bigl(C(t)+1-\rho^{I}(t)-\rho^{I\!I}(t)\bigr) \quad&
 \quad \rho_{e*}(t) = \frac{1}{2}\bigl(1-C(t)-\rho^{I}(t)+\rho^{I\!I}(t)\bigr)\\ \\
  \rho_{*e}(t) = \frac{1}{2}\bigl(1-C(t)+\rho^{I}(t)-\rho^{I\!I}(t)\bigr) \quad&
 \quad \rho_{**}(t) = \frac{1}{2}\bigl(C(t)-1+\rho^{I}(t)+\rho^{I\!I}(t)\bigr)
 \, .
  \end{array}
\end{equation}
These equations allow for a calculation of the initial values for
$\rho_{ee}$, $\rho_{e*}$, $\rho_{*e}$ and $\rho_{**}$ from the given
initial conditions $C(t=0)=C$, $\rho^{I}(t=0)=\rho^{I}$ and
$\rho^{I\!I}(t=0)=\rho^{I\!I}$. The approximation for the time-flow of
the observables is obtained from Eqs.~\eqref{gl_corrapprox} through
numerical integration. The correlated-$\rho$-approximation for $C(t)$
is then obtained via Eq.~\eqref{gl_zshgcorrrho0011}. We add that the
approximation was derived following the lines of PAS.

\subsubsection*{Stability of the uncorrelated configurations}

After the more general calculation for two correlated systems we
return to the study of the autocorrelation function of a single
system. This can be easily achieved by setting
$q_0^{I}=q_1^{I}=q_s^{I}=0$ and $q_0^{I\!I}=q_0$, $q_1^{I\!I}=q_1$ and
$q_s^{I\!I}=q_s$ in Eqs.~\eqref{gl_corrapprox}, i.e. we freeze system
$I$ in its initial configuration which we assure to be identical to
that of system $I\!I$ by imposing $C(t=0)=1$.

We mention that, even though $C$ is a dynamical quantity, the subsequent
study does not primarily deal with strictly speaking non-equilibrium
systems. The interest will be shifted to non-equilibrium
situations of the correlation between systems that are themselves in
equilibrium. Consequently, we assume that system $I\!I$ has a constant
particle density, i.e. $\rho^{I} = \rho^{I\!I} = \rho =
\text{const}$. Under these premises Eqs.~\eqref{gl_zshgCrho} become
\begin{equation} \label{gl_zshgCrhoauto}
\begin{array}{c}
  \rho_{ee}(t) = \frac{1}{2}\bigl(C(t)+1-2\rho\bigr) 
 \qquad \rho_{**}(t) = \frac{1}{2}\bigl(C(t)-1+2\rho\bigr)\\ \\
 \quad \rho_{e*}(t) =  \rho_{*e}(t) = \frac{1}{2}\bigl(1-C(t)\bigr)
 \, .
\end{array}
\end{equation}

Using this result, we obtain a closed differential equation $\dot{C} =
\dot{C}(C)$ for $C(t)$ when we sum up Eqs.~\eqref{gl_corrapprox00} and
\eqref{gl_corrapprox11}. One can show that $C_{\text{li}} =
\rho^2+(1-\rho)^2$, which corresponds to uncorrelated configurations,
is a fixed point of the differential equation for $C(t)$. In order to
check its stability, we calculate
\begin{equation}
 \frac{\partial\dot{C}}{\partial C}\bigg|_{C=C_{\text{li}}} =
 \frac{\bigl(1-(p+1)\rho\bigr)^k\bigl((\sqrt{kp\,}-1)\rho+1\bigr)
 \bigl(q_1(1-\rho)(1-(p+1)\rho)+q_s\rho\bigr)}{\rho(1-\rho)^{k+2}
 \bigl(p\rho^2+(1-\rho)^2\bigr)} \, \times \,
 \bigl((\sqrt{kp\,}+1)\rho-1\bigr) \, .
\label{eq:crelax}
\end{equation}
Here $\rho$ determines the chemical potential $\mu$ via the solution
for the liquid phase and $q_0=q_1/e^{\mu}$ is given through detailed
balance.

The fraction on the r.h.s. is always positive, the sign of the
expression only depends on the last factor. We conclude that the fixed
point $C_{\text{li}}$ is stable for
\begin{equation}
  \rho < \frac{1}{\sqrt{kp\,}+1} \equiv \rho_g \ ,
\end{equation}
and unstable for $\rho>\rho_g$. Here $\rho_g$ coincides with the
spin-glass instability calculated in \cite{HaWe} which is the local
instability of the liquid solution toward a 1RSB, i.e.~glassy,
solution. In the case of a continuous transition ($p=1$), $\rho_g$ is
the transition density from the liquid to the spin-glass phase. The result
appears very reasonable, as we expect the system to become,
physically speaking, very ``viscous'' when the density approaches
$\rho_g$ coming from the liquid phase. This should result in a
divergence of the relaxation time of $C(t)$ to $C_{\text{li}}$ which
is implied in the stability result for the fixed point. We shall
compare the results for the relaxation time with the values extracted
from MC simulations in Sec.~\ref{sec:MCglass}.

In the case of a discontinuous transition ($p>1$) the situation is
more complex (see \cite{weigt,HaWe}). The relaxational dynamics is
governed by metastable glassy states that appear for densities {\em
below} the spin-glass instability at $\rho_g$. The result for the
stability of $C_{\text{li}}$ is unphysical then as the divergence of
the relaxation time should occur at lower densities (depending on
initial conditions etc.). This inconsistency is, however, not
surprising as we can show by a 1RSB treatment of the ensemble given by
$\Xi_G$ with the observables $\rho_{ee}$, $\rho_{e*}$, $\rho_{*e}$ and
$\rho_{**}$ that, in the discontinuous case ($p>1$), the vicinity of
$C_{\text{li}}$ is not described correctly by a replica symmetric
approach (which is the approach that was used for the
correlated-$\rho$-approximation) when the density is sufficiently
close to $\rho_g$. We shall therefore only consider the continuous
glass-transition ($p=1$), i.e. the ordinary Bethe lattice, in the
following.

\subsection{Comparison with MC simulations}
\label{sec:MCglass}

We have performed MC simulations for a lattice ($k=2$, $p=1$) with $N
= 5\cdot10^6$ sites which was realised as a regular random graph. A
given dynamics $(q_0, q_1, q_s)$ was performed for a time $T/2$ in
order to equilibrate the system. Subsequently the autocorrelation
function $C(t)$ was recorded w.r.t. $t_1 = T/2$ for another time
interval of $T/2$. In order to guarantee equilibration in the first
half of the simulation time, we have to choose $T\gg \tau_{eq}$. To
achieve this, we have fixed $T$ such that the decay of $C(t)$ to
$C_{\text{li}}$ could be observed. The dynamics was performed without
particle movement, i.e. $q_0=\min(1,e^{-\mu})$, $q_1=\min(1,e^{\mu})$,
$q_s=0$.

\subsubsection*{Asymptotic relaxation time}

\begin{figure}[tbp]
  \begin{center}
    \includegraphics[height = 7cm]{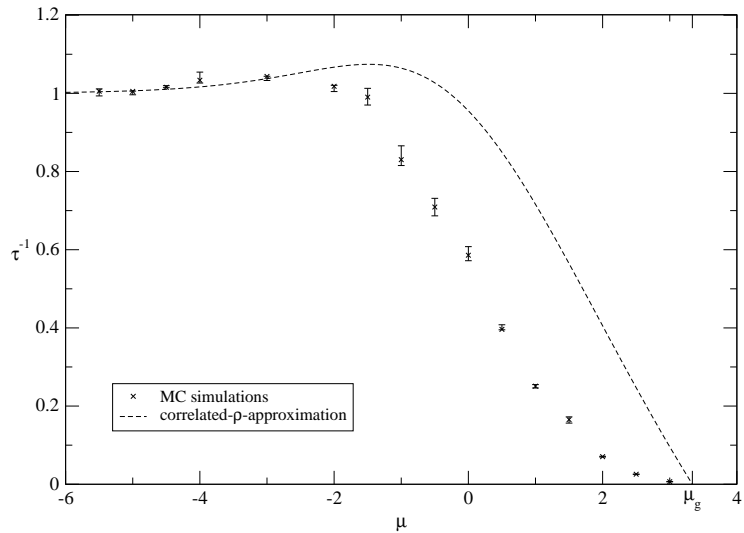}
    \caption{The inverse asymptotic relaxation time of $C(t)$ toward
      $C_{\text{li}}$ from numerical simulations (upper bounds)
      compared with the relaxation time from the
      correlated-$\rho$-approximation.}
    \label{fig_Crelax} 
  \end{center}
\end{figure}

From the decay curves of $C(t)$ we can extract asymptotic relaxation
times through a linearization of the function $f(t) = \ln
(C(t)-C_{\text{li}})$ near $t \simeq T$. From the simulations we
obtain $f(t)$ as a convex function, which means that the extracted
relaxation times must be considered as lower bounds for the actual
relaxation times. Results for the inverse asymptotic relaxation time
$\tau^{-1}$ are shown in Fig.~\ref{fig_Crelax}. The numerical data are
plotted together with the result of the
correlated-$\rho$-approximation, which follows directly from
Eq.~(\ref{eq:crelax}).

As mentioned before, the approximation correctly predicts a divergence
of the equilibration time at $\mu_g\simeq 3.34$. The quantitative agreement is,
however, poor when we approach the phase transition, only fare away
numerical and approximate analytical data are close to each other. The
figure indeed indicates that also the critical exponent of the
equilibration-time divergence is different for the two cases. This
suggests that the correlated $\rho$-approximation is still missing the
asymptotically dominating slow process, i.e.~a more detailed and well
chosen set of order parameters is needed.

\subsubsection*{Decay of the autocorrelation function}

In order to complete the picture, we show the decay curves of $C(t)$
for different $\mu$. The time axis has been shifted as to map $T/2
\to 0$.

\begin{figure}[htbp]
  \begin{center}   
    \includegraphics[height = 8cm]{corrNeg.eps} \qquad
    \includegraphics[height = 8cm]{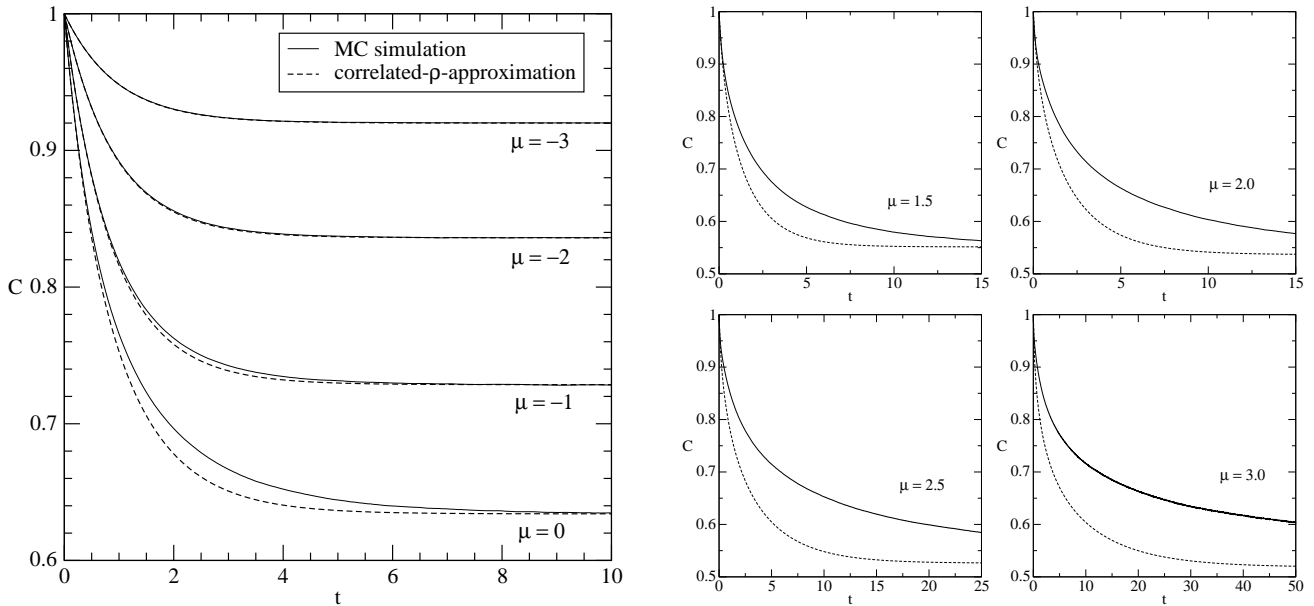}
    \caption{Decay curves of the autocorrelation function $C(t)$ for
   various $\mu$.}
    \label{fig_corr} 
  \end{center}
\end{figure}

The left of Figs.~\ref{fig_corr} shows the curves for small $\mu$,
i.e.~far away from the spin-glass transition point. The qualitative
and quantitative agreement between numerical and analytical data is
quite satisfactory, even if the differences grow with larger values of
the chemical potential. Note in particular that the approximation
results are very good for short times, where by construction the
equipartition assumption of PAS is valid, and for large times, where
again by construction the assumption becomes correct. The largest
deviation is found for intermediate times.

The right of the figures shows analogous curves for positive
$\mu$. Since we are coming closer to the transition, the deviations
show up more drastically, and the quantitative agreement for
intermediate times becomes quite poor. Still, the initial part and
the asymptotic value are reproduced correctly by the
correlated-$\rho$-approximation.

\section{Conclusion and outlook}

In this paper, we have discussed the dynamical behaviour of a
hard-sphere lattice-gas model on generalised Bethe lattices. The paper
can be seen as a natural extension of the results presented in a
preceding publication \cite{HaWe} which was entirely dedicated to the
study of equilibrium properties.

Since a technique allowing for an exact analysis of the
non-equilibrium dynamics of finite-connectivity systems is still
missing, we have applied a projective approximation scheme. Within
this scheme, the dynamics is approximated by the time evolution of a
finite number of global observables. Since the exact equations for
their time evolution do not close, we have to use an approximate
closure scheme. We have argued that the best and most natural closure
scheme based only on the considered variables is the following: The
true non-equilibrium state of the system is approximated by a
pseudo-equilibrium distribution which is flat over all microscopic
configurations being consistent with the values of all considered
observables. The quality of this approximation depends crucially on
the selection of the observable set: The latter has to include for
sure a set of observables which are able to describe correctly the
equilibrium state and to identify different thermodynamic phases. In
general, a larger observable set gives also quantitatively better
results, but the complexity of the underlying calculations increases
considerably.

As a first application, we have studied the dynamics close to the
liquid-crystal transition, both in the case of a second- and a
first-order transition. Already the minimal description shows a very
good qualitative and reasonable quantitative agreement with
Monte-Carlo simulations, a more sophisticated level of approximation
is hardly distinguishable from the numerical findings.

As a second application, we have also studied the dynamics close to
the continuous spin-glass transition for the model on ordinary Bethe
lattices. The chosen minimal approach is able to reproduce correctly
the phase transition point, where the relaxation time diverges, but
the quantitative agreement between numerical and analytical data for
the dynamical behaviour close to the phase transition is quite
poor. This signals that the selected observable set is not yet able to
include the slowest relaxation process, and richer observable sets
have to be used. We have in fact tried to do so, but the complexity of
the resulting calculations was too large to arrive at a satisfactory
result. Concerning the approximation scheme close to the spin glass
transition, one probably should also go to observable sets which allow
for the inclusion of explicitly replica symmetry breaking
fluctuations. It may well be that the presented approach is able to
reflect the longitudinal instability of the liquid solution within the
replica-symmetric framework, and not the important replicon
fluctuations breaking the replica symmetry \cite{MePaVi}. Further
research is necessary in this direction.

Another open question concerns the divergence of the equilibration
time close to a discontinuous spin-glass transition as occurring on the
generalised Bethe lattices. The latter is of fundamental interest in
the context of the structural glass transition, but cannot be seen
using PAS for the simple observable sets considered so far. This would
also be of importance for the applicability of the method to
kinetically constrained models \cite{RiSo,SeBiTo}, which have a trivial
thermodynamics, but show a dynamical arrest due to kinetic
constraints.

{\bf Acknowledgement:} We are grateful to G. Semerjian and A. Zippelius
for many interesting discussions. HHG acknowledges also the
hospitality of the ISI Foundation in Turin, where parts of this
research were worked out.

\appendix
\section{Contributions to the changes in $\sigma_j$}

\label{app_contrisigma}

We introduce the following functions of $j$:
\begin{eqnarray}
  P_0^{(0/1)}(j) & = & 
  \begin{cases} \frac{\frac{k+1-j}{k+1}\,\sigma_j^{(0/1)}}
    {\sum_{l=0}^{k} \frac{k+1-l}{k+1}\,\sigma_l^{(0/1)}} \quad \text{for } j
    = 0, \ldots , k\\
    0 \quad \text{otherwise} 
  \end{cases}
\\
  P_1^{(0/1)}(j) & = & 
  \begin{cases} \frac{\frac{j+1}{k+1}\,\sigma_{j+1}^{(0/1)}}
    {\sum_{l=0}^{k} \frac{l+1}{k+1}\,\sigma_{l+1}^{(0/1)}} \quad \text{for }
    j = 0, \ldots , k\\
    0 \quad \text{otherwise}
  \end{cases}
\end{eqnarray}

$P_0^{(0/1)}(j)$ gives the probability that a 0/1-site, which is
contained in an empty clique, has $j$ occupied neighbouring
sites. $P_1^{(0/1)}(j)$ gives the probability that a 0/1-site, which is
empty and which is contained in a clique that carries exactly one
particle, has $j$ more occupied neighbouring sites (in other adjacent
cliques). The values are exact in the $\Xi_G$-ensemble.

The actions of the dynamics are listed in the following together with their
probabilities to occur: \\

\begin{tabular}{|c|l|l|}
\hline
ID & Action & Probability \\
\hline
$\mathcal{R}0$ & Removal of a particle from the 0-lattice & $\frac{p}{p+1}\,\rho^{(0)}\,q_0$\\
$\mathcal{I}0$ & Insertion of a particle to the 0-lattice & $\frac{p}{p+1}\,\sigma_0^{(0)}\,q_1$\\
$\mathcal{R}1$ & Removal of a particle from the 1-lattice & $\frac{1}{p+1}\,\rho^{(1)}\,q_0$\\
$\mathcal{I}1$ & Insertion of a particle to the 1-lattice & $\frac{1}{p+1}\,\sigma_0^{(1)}\,q_1$\\
$\mathcal{J}00$& Jump of a particle from a 0-site to a 0-site &
$\frac{p}{p+1}\,\rho^{(0)}\,q_s\,\frac{p-1}{p}\,P_1^{(0)}(0)$\\
$\mathcal{J}01$& Jump of a particle from a 0-site to a 1-site &
$\frac{p}{p+1}\,\rho^{(0)}\,q_s\,\frac{1}{p}\,P_1^{(1)}(0)$\\
$\mathcal{J}10$& Jump of a particle from a 1-site to a 0-site &
$\frac{1}{p+1}\,\rho^{(1)}\,q_s\,P_1^{(0)}(0)$\\
\hline
\end{tabular} 
\newpage
The following table contains the contributions of the different
actions of the dynamics to the changes in $\sigma_j^{(0/1)}$ or more
precisely to the changes in the corresponding numbers of
vacancies. The contributions are averages which are exact in the
$\Xi_G$-ensemble apart from the exact $\delta$-contributions. \\

\begin{tabular}{|c|p{5cm}|p{5cm}|c|}
\hline
ID & Contribution to $\sigma_j^{(0)}$ & Contribution to
$\sigma_j^{(1)}$ & Nature of the contribution\\
\hline\hline
 $\mathcal{R}0$ 
    & $+\,\delta_{0,j}$ &  & direct \\
    & $+\,(k+1)\,(p-1)\,P_1^{(0)}(j)$ & $+\,(k+1)\,P_1^{(1)}(j)$ & neighbours\\
    & $-\,(k+1)\,(p-1)\,P_1^{(0)}(j-1)$ & $-\,(k+1)\,P_1^{(1)}(j-1)$ &
    neighbours \\
\hline
 $\mathcal{I}0$ 
    & $-\,\delta_{0,j}$ &  & direct \\
    & $+\,(k+1)\,(p-1)\,P_0^{(0)}(j-1)$ & $+\,(k+1)\,P_0^{(1)}(j-1)$ &
    neighbours \\
    & $-\,(k+1)\,(p-1)\,P_0^{(0)}(j)$ &  $-\,(k+1)\,P_0^{(1)}(j)$ &
    neighbours \\
\hline
 $\mathcal{R}1$
    &  & $+\,\delta_{0,j}$ & direct \\
    & $+\,(k+1)\,p\,P_1^{(0)}(j)$ &  & neighbours \\
    & $-\,(k+1)\,p\,P_1^{(0)}(j-1)$ &  & neighbours \\
\hline
 $\mathcal{I}1$
    &  & $-\,\delta_{0,j}$ & direct \\
    & $+\,(k+1)\,p\,P_0^{(0)}(j-1)$ & & neighbours \\
    & $-\,(k+1)\,p\,P_0^{(0)}(j)$ & & neighbours \\
\hline
 $\mathcal{J}00$
    & $+\,k\,(p-1)\,P_1^{(0)}(j)$ & $+\,k\,P_1^{(1)}(j)$ & backward \\
    & $-\,k\,(p-1)\,P_1^{(0)}(j-1)$ & $-\,k\,P_1^{(1)}(j-1)$ & backward \\
    & $+\,k\,(p-1)\,P_0^{(0)}(j-1)$ & $+\,k\,P_0^{(1)}(j-1)$ & forward \\
    & $-\,k\,(p-1)\,P_0^{(0)}(j)$ & $-\,k\,P_0^{(1)}(j)$ & forward \\
\hline
 $\mathcal{J}01$
    & $+\,\delta_{1,j}$ & $-\,\delta_{1,j}$ & intra-clique \\
    & $+\,k\,(p-1)\,P_1^{(0)}(j)$ & $+\,k\,P_1^{(1)}(j)$ & backward \\
    & $-\,k\,(p-1)\,P_1^{(0)}(j-1)$ & $-\,k\,P_1^{(1)}(j-1)$ & backward \\
    & $+\,k\,p\,P_0^{(0)}(j-1)$ & & forward \\
    & $-\,k\,p\,P_0^{(0)}(j)$ & & forward \\
\hline
 $\mathcal{J}10$
    & $-\,\delta_{1,j}$ & $+\,\delta_{1,j}$ & intra-clique \\
    & $+\,k\,p\,P_1^{(0)}(j)$ & & backward  \\
    & $-\,k\,p\,P_1^{(0)}(j-1)$ & & backward \\
    & $+\,k\,(p-1)\,P_0^{(0)}(j-1)$ & $+\,k\,P_0^{(1)}(j-1)$ & forward \\
    & $-\,k\,(p-1)\,P_0^{(0)}(j)$ & $-\,k\,P_0^{(1)}(j)$ & forward \\
\hline
\end{tabular}
\\ \\ \\
The differential equations for the time-evolution of the observables
$\sigma_j^{(0/1)}$ are obtained as follows. Let $N_j^{(0/1)}$ be
the number of vacancies corresponding to $\sigma_j^{(0/1)}$. The
average change of $N_j^{(0/1)}$ during the Monte-Carlo-time $\Delta t
= 1/N$ be $\Delta N_j^{(0/1)}$ which is obtained by choosing the
contributions from the respective column in the above table,
multiplying them with the probabilities for the corresponding actions of
the dynamics and summing up the terms. The equations for $\Delta
N_j^{(0/1)}$ become differential equations for $\sigma_j^{(0/1)}$ in
the thermodynamic limit ($N \to \infty$) according to
\begin{equation}
  \Delta N_j^{(0/1)} = \frac{N^{(0/1)}} {N} \:
  \frac{\Delta(N_j^{(0/1)}/N^{(0/1)})}{\Delta t} = 
   \begin{cases}
     \:\frac{p}{p+1} \,\dot{\sigma}_j^{(0)} \\
     \:\frac{1}{p+1} \,\dot{\sigma}_j^{(1)}
   \end{cases}
\end{equation}

Here $N^{(0/1)}$ is the number of sites belonging to the respective sublattice.

\end{document}